\pdfoutput=1
\documentclass[final]{amsart}

\usepackage{amssymb}

\usepackage{graphics} 
\usepackage{latexsym}
\usepackage{epsfig}
\usepackage{subfigure}
\usepackage{float}

\graphicspath{{./figures/}}

\newcommand{\R}{{\mathbb R}}

\newcommand{\J}{{\bf J}}
\newcommand{\E}{{\bf E}}

\newcommand{\B}{{\bf B}}
\newcommand{\bj}{{\bf j}}

\newcommand{\be}{{\bf e}}
\newcommand{\bq}{{\bf q}}

\title[Conditionally Gaussian Hypermodels for Cerebral
Sources]{Conditionally Gaussian Hypermodels for Cerebral Source
Localization}

\author{Daniela Calvetti \and Harri Hakula \and Sampsa Pursiainen \and Erkki Somersalo}
\address{D. ~Calvetti:\ Department of Mathematics, Case Western Reserve University, 10900 Euclid
Avenue, Cleveland, OH 44106, USA} \email{daniela.calvetti@case.edu}
\address{H. ~Hakula:\ Institute of Mathematics, Box 1100, FI-02015
Helsinki University of Technology, Finland}
\email{harri.hakula@tkk.fi} \address{S. ~Pursiainen:\ Institute of
Mathematics, Box 1100, FI-02015 Helsinki University of Technology,
Finland} \email{sampsa.pursiainen@tkk.fi}
\address{E. ~Somersalo:\ Department of Mathematics,
Case Western Reserve University, 10900 Euclid Avenue, Cleveland, OH
44106, USA} \email{erkki.somersalo@case.edu}

\begin{document}

\begin{abstract}
Bayesian modeling and analysis of the MEG and EEG modalities provide
a flexible framework for introducing prior information complementary
to the measured data. This prior information is often qualitative in
nature, making the translation of the available information into a
computational model a challenging task. We propose a generalized
gamma family of hyperpriors which allows the impressed currents to
be focal and we advocate a fast and efficient iterative algorithm,
the Iterative Alternating Sequential (IAS) algorithm for computing
maximum a posteriori (MAP) estimates. Furthermore, we show that for
particular choices of the scalar parameters specifying the
hyperprior, the algorithm effectively approximates popular
regularization strategies such as the Minimum Current Estimate and
the Minimum Support Estimate. The connection between
priorconditioning and adaptive regularization methods is also
pointed out. The posterior densities are explored by means of a
Markov Chain Monte Carlo (MCMC) strategy suitable for this family of
hypermodels. The computed experiments suggest that the known
preference of regularization methods for superficial sources over
deep sources is a property of the MAP estimators only, and that
estimation of the posterior mean in the hierarchical model is better
adapted for localizing deep sources.
\end{abstract}


\maketitle

\section{Introduction}

The human brain contains approximately $10^{11}$ exitable neurons
whose resting state is characterized by a cross-membrane voltage
difference. Electromagnetic signals propagate as perturbations of
this voltage difference, or action potentials, along the axons and
are transferred across the synaptic gaps by neurotransmitters,
creating a post-synaptic potential in the receiving neurons. The
post-synaptic potential may be relatively stable over a period of
milliseconds and, being well localized, it can be modelled
mathematically as current dipole. The neurons are organized in
bundles, and when thousands of neighboring neurons are
simultaneously in the post-synaptic excitation state, the net effect
of the post-synaptic potentials gives rise to a localized current
approximately parallel to the neuron bundle. This elementary
impressed source current drives an Ohmic volume current in the brain
tissue, and the net electromagnetic field can be registered on or
outside the skull. Imaging of the neuronal activity based on the
registered electric voltage (EEG) or on the magnetic field (MEG) has
become a standard research tool in clinical and cognitive studies.
Furthermore, when coupled to functional imaging methods, MEG and EEG
have a great potential to gather pertinent information about the
coupling between neuronal activity and cerebral hemodynamics.

The advantage of electromagnetic brain imaging modalities is their
good temporal resolution, about a millisecond, while their spatial
resolution is limited by various factors, including weak
signal-to-noise ratio, ambiguities in the source estimation due to
the non-uniqueness of the inverse source problem \cite{sarvas} and
uncertainties in the model. For example, while EEG suffers from lack
of knowledge of the electric conductivity distribution and
anisotropy of the head, MEG is known to be less affected by that
\cite{gencer,wolters}. For both modalities, the anisotropy of the
white matter may result in a strong bias of the volume currents, an
effect that should be taken into account in a detailed model. 

The properties of the solutions to electromagnetic inverse source
problems depend on the {\em a priori} information implemented in the
algorithm. One piece of information that may certainly improve the
spatial resolution, if properly incorporated in the algorithm, is
the focal nature of the impressed currents. In fact, since the
electric and magnetic fields outside the skull depend linearly on
the impressed currents, but the component of the source belonging to
the significantly nontrivial null space of the forward map has no
effect on the data, the determination of a physiologically
meaningful solution must be based on complementary information. The
implementation of feasible selection criteria has led to different
solutions. For example, setting the null space component to zero
gives the {\em minimum norm estimate} (MNE) \cite{ilmoniemi}, while
minimization of the current, or $\ell^1$ norm of the source results
in the {\em minimum current estimate} (MCE) \cite{uutela}. Solutions
such as the Low Resolution Electromagnetic Tomography (LORETA)
\cite{pascual-marqui} are based on the assumption of smoothness of
the source. In \cite{jerbi}, the localization is pursued by local
multipole expansions of the fields. Low-dimensional parametric
models, for example those using only a few current dipoles
\cite{demunck,mosher,hamalainen}, lead to localized solutions but
limit the number of possible source configurations. Prior anatomical
information such as the location of the sulci has been shown to
improve the performance of the localization \cite{baillet0}.

Bayesian methods are widely used in EEG/MEG to implement pertinent
prior information such as anatomic constraints or functional
information based on other imaging modalities
\cite{baillet0,jun,schmidt,trujillo-barreto}. An additional level of
flexibility to Bayesian modelling is provided by hierarchical models
that allow uncertainties in the prior model itself
\cite{auranen,nummenmaa,sato}. In particular, the model hierarchy is
a powerful tool for including prior information that is qualitative
rather than quantitative \cite{CS_IP2}.

The connection between classical regularization methods and Bayesian
Maximum A Posteriori (MAP) estimates is well known. In particular,
the Gaussian models lead to a MAP estimate that coincides with the
standard Tikhonov regularized solution with a quadratic penalty,
while non-Gaussian models are needed for non-quadratic penalties
that are often more useful but computationally more challenging. In
this work, we construct a {\em conditionally} Gaussian hierarchical
parametric model that has the computational advantages of Gaussian
prior models but leads to a rich class of MAP estimators that have
the desirable qualitative properties of numerous commonly used
non-Gaussian models. Using the conditional normality, we construct a
fast, efficient and simple MAP estimation algorithm, and show that
with proper choices of the few model parameters, the algorithm can
be interpreted as a fixed point iteration for solving the minimum
norm estimate, the minimum current estimate and, more generally, the
minimum $\ell^p$ estimate, while with a different choice, the
algorithm approximates the {\em minimum support estimate} (MSE)
\cite{nagarajan}. Hence, our approach puts these methods in a
unified framework.

The estimation algorithms require the solution of large linear
systems, the system size depending on the discretization of the
model. When a realistic three-dimensional model is used as in the
present paper, iterative linear systems solvers are indispensable.
It is common practice to improve the performance of the iterative
solvers by preconditioners \cite{saad}. Recently, the authors
introduced the concept of {\em priorconditioner} \cite{CS}, a
preconditioner that is based on the prior model rather than on
linear algebraic properties of the system matrix. It is shown in
this article that the priorconditioning based on the conditionally
Gaussian hierarchical models yields an effective implementation of
adaptive regularization similar to the FOCUSS (FOCal Underdetermined
System Solver) algorithm \cite{gorodnitsky}.

A well-known feature of many regularization based source
localization algorithms, including the minimum norm and minimum
current estimates, is their tendency to favor surface sources,
leading sometimes to a gross misplacement of deep sources. To
suppress this bias, different weighting methods have been proposed
to favor deep sources over the shallow ones, see, e.g., \cite{lin}
for a recent account. From the statistical point of view, however, a
depth dependent prior covariance which compensates for the
preference of the likelihood towards superficial sources appears
rather dubious, since there is no reason to believe a priori that a
deep source should have higher variance than a superficial one. In
this article, we confirm with numerical experiments that surface
biasing may be a property of the MAP estimate, while Markov Chain
Monte Carlo (MCMC) sampling based posterior mean estimates may
localize better deep sources without requiring weight compensation.
This result reinforces the concept, well known to the Bayesian
statistics community, that the mode of the posterior distribution
may do a poor job at representing the distribution, demonstrating
that the Bayesian formulation of the inverse problem is much more
than ``yet another way to regularize an ill-posed problem'', as is
sometimes incorrectly stated.

Preliminary testing of the inversion algorithms and the MCMC
sampling is done using a simple geometry with constant conductivity,
modelling the distributed currents by a field of current dipoles.
Subsequently, numerical experiments with a realistic three
dimensional head model, with different electric conductivities in
the scalp, skull, cerebrospinal fluid and the brain tissue are also
presented. The volume current calculations are carried out with a
finite element algorithm developed for this purpose. The distributed
currents are represented by Raviart-Thomas elements that are a
reasonable substitute for singular dipoles in the FEM context.

\section{Forward model, EEG and MEG}

We start by introducing the notations to be used in the sequel and
review some basic facts concerning the forward model, which is based
on the standard approach using the quasi-static approximation of
Maxwell's equations \cite{sarvas}. We denote by $D\subset\R^3$ the
head with boundary surface $S$ and scalar conductivity distribution
$\sigma>0$. If we let $\J$ denote the impressed current density in
$D$, the total current density, consisting of the impressed current
and the Ohmic volume current, is $\J_{\rm tot} = \J +\sigma\E$,
where $\E$ is the electric field induced by the current. Under the
quasi-static approximation, we may assume that $\E$ is conservative,
$\E=-\nabla u$. Neglecting the electric displacement current in the
Maxwell-Amp\`{e}re equation, the total current density is divergence
free, leading to the Poisson equation for the electric potential,
\begin{equation}\label{poisson}
 \nabla\cdot(\sigma \nabla u)=\nabla\cdot \J,\quad \frac{\partial
 u}{\partial n}\bigg|_S=0,
\end{equation}
where the Neumann boundary condition follows from the assumption
that the conductivity vanishes outside $D$. The magnetic field $\B$
outside the head induced by the total current is, according to the
Biot-Savart law, obtained as
\begin{equation}\label{biot savart}
\B(x) =\frac{\mu_0}{4\pi}\int_D \J_{\rm tot}\times
\frac{x-y}{|x-y|^3} dy,\quad x\in\R^3\setminus D.
\end{equation}
The computation of the magnetic field therefore consists of two
steps: the numerical solution of the Poisson equation
(\ref{poisson}) to find $u$ and thus the total electric current
density $\J_{\rm tot}=\J-\sigma\nabla u$, and the computation of the
integral yielding the magnetic field $\B$ (\ref{biot savart}). Each
of these steps poses computational challenges.

The solution of the boundary value problem (\ref{poisson}) can be
formally expressed in terms of the Neumann Green's function
${\mathcal G}_{\rm N}$ of the diffusion operator
$\nabla\cdot\sigma\nabla$,
\begin{equation}\label{integral for u}
 u(x)=\int_D {\mathcal G}_{\rm N}(x,y)\nabla\cdot \J(y) dy.
\end{equation}
We assume that the electric potential is measured at locations
$x_\ell$, $1\leq\ell \leq L$ on the surface $S$ of the head. The
approximation of the impressed current by a finite linear
combination $\J =\sum_{k=1}^K \alpha_k\bj_k$ of basis currents
$\bj_k$, $1\leq k\leq K$, leads to a discrete model
\begin{eqnarray*}
 u_\ell &=& u(x_\ell) = \sum_{k=1}^K\left(\int_D
 {\mathcal G}_{\rm N}(x_\ell,y)
 \nabla\cdot \bj_k(y) dy\right)\alpha_k\\
 \noalign{\vskip4pt}
 &=& \sum_{k=1}^K M_{\ell k}^{\rm e} \alpha_k,\quad 1\leq\ell\leq
 L,
\end{eqnarray*}
where the matrix $M^{\rm e}\in\R^{L\times K}$ is the {\em electric
lead field matrix}. Similarly, assume that outside the head at
points $x_n$, $1\leq n\leq N$, the projection of the magnetic field
in given directions $\be_n$ is measured. Substituting the expression
(\ref{integral for u}) in the Biot-Savart law and representing the
impressed current in the basis $\bj_k$, we have the discrete model
\begin{eqnarray*}
v_n &=& \be_n\cdot\B(x_n)\\
&=&\sum_{k=1}^K \alpha_k  \be_n \frac {\mu_0}{4\pi} \int_D
\left\{\bj_k(y) -\sigma(y)\nabla\int_D{\mathcal G}_{\rm
N}(y,z)\nabla\cdot\bj_k(z)dz\right\}\! \times \!
\frac{x_n-y}{|x_n-y|^3}
dy \\
\noalign{\vskip4pt} &=& \sum_{k=1}^K M_{nk}^{\rm m}\alpha_k,\quad
1\leq n\leq N,
\end{eqnarray*}
where the matrix $M^{\rm m}\in\R^{N\times K}$ is the {\em magnetic
lead field matrix}.

The EEG inverse source problem is to estimate the vector
$\alpha\in\R^K$ from the noisy observations of the voltage potential
$u\in\R^L$, while in the MEG inverse source problem the data consist
of the noisy observations of the magnetic field component vector
$v\in\R^N$.

In our model, we assume that the current elements $\bj_k$ that
constitute the basis for the distributed current are either dipoles
or dipole-like vector-valued elements.

\section{Imaging in the Bayesian framework}

In the Bayesian framework, inverse problems are recast in the form
of statistical inference \cite{CS,KS}. The lack of information about
any of the quantities appearing in the formulation of the problem is
expressed by modelling them as random variables, and the available
information is encoded in the probability densities.

In the electromagnetic inverse source problem, the goal is to
estimate the coefficient vector $\alpha$ from the observations
\[
b = M \alpha + e,
\]
where $M$ is either the electric or magnetic lead field matrix and
$e$ is noise that, for simplicity, is modelled here as additive.
Although not necessary, we assume that the noise is white Gaussian
with known variance $\sigma^2$, which we assume known, leading to a
likelihood model of the form
\[
 \pi(b\mid \alpha) \propto{\rm exp}\left(-\frac 1{2\sigma^2}\|
 b-M\alpha\|^2\right).
\]
We consider prior models that are conditionally Gaussian and of the
form
\[
 \pi_{\rm prior}(\alpha \mid\theta)\propto{\rm exp}\left(-\frac 12\|
 D^{-1/2}_\theta \alpha\|^2
 -\frac12\sum_{j=1}^K\log\theta_j\right),
\]
where $D_\theta$ is a diagonal matrix, $D_\theta = {\rm
diag}(\theta_1,\ldots,\theta_K)$, and the logarithmic term comes
from normalizing of the prior density by the determinant of
$D^{-1/2}_\theta$. The posterior density conditional on $\theta$ is,
by Bayes' formula,
\begin{eqnarray*}
 \pi(\alpha\mid b,\theta) & \propto & \pi_{\rm
 prior}(\alpha\mid\theta)\pi(b\mid \alpha) \\ & \propto & {\rm exp}
\left(-\frac 1{2\sigma^2}\|
 b-M\alpha\|^2-\frac 12\|
 D^{-1/2}_\theta \alpha\|^2 -\frac12\sum_{j=1}^K\log\theta_j\right).
\end{eqnarray*}
Assuming the variance vector $\theta$ known and fixed, the MAP
estimate for $\alpha$,
\[
 \alpha_{\rm MAP} = {\rm argmin}\left(\frac 1{2\sigma^2}\|
 b-M\alpha\|^2+\frac 12\|
 D^{-1/2}_\theta \alpha\|^2\right),
\]
is the classical Tikhonov regularized solution with a penalty
defined by the diagonal matrix $D$. It is known that if $\theta$ has
equal entries, this solution is smeared out even if the data
corresponds to a focal input. To improve the localization,
non-quadratic penalties, for example the $\ell^p$-norm, $p< 2$, of
the coefficient vector $\alpha$, have been proposed. Here, we take a
different approach assuming instead that the variance vector
$\theta$ is unknown, and thus making its estimation a part of the
inverse problem. The variance vector $\theta$ is modelled as a
random variable, and available a priori information concerning it is
expressed by a hyperprior $\pi_{\rm hyper}(\theta)$. The prior
probability density of the pair $(\alpha,\theta)$ is then
\[
 \pi_{\rm prior}(\alpha,\theta)=\pi_{\rm hyper}(\theta)\pi_{\rm
 prior}(\alpha\mid\theta),
\]
and, according to Bayes' formula, the posterior probability density,
conditioned on the observation $b$ alone, becomes
\[
 \pi(\alpha,\theta\mid b)\propto\pi_{\rm hyper}(\theta)\pi_{\rm
 prior}(\alpha\mid\theta)\pi(b\mid\alpha).
\]
This implies that in the present formulation we need to estimate
both $\alpha$ and its prior variance vector $\theta$.

\section{Hypermodels: MCE, minimum $\ell^p$  and beyond}

Prior densities require quantitative information about the unknown,
e.g., an estimate for its mean and its dynamical range, which is in
turn related to the prior variance. The flexibility of hypermodels
lie in their ability to import {\em qualitative} information into
the estimation process, see \cite{CS_IP1,CS_IP2,CS_SPIE,CS}. In this
article, we assume that the only a priori information concerning the
impressed current is that it should consist of few focal sources. In
statistical terms, such information can be expressed by the
following three statements:
\begin{enumerate}
\item Nearby source current elements, should not be, a priori,
mutually dependent, to favor the focality; \item No location
preference for the activity should be given a priori; \item Most of
the dipole-like sources should be silent, while few of them could
have large amplitude.
\end{enumerate}
To encode these conditions into the hyperprior, we observe first
that stochastic dependence among the variances $\theta_k$ would
couple the dynamical ranges of the corresponding coefficients
$\alpha_k$. Therefore, it is reasonable to assume that the variances
$\theta_k$ are mutually independent. On the other hand, without
prior knowledge about the location of the active sources, it is also
reasonable to assume a priori that the variances are equally
distributed. Furthermore, we want to allow the distribution of the
variances to favor small values while permitting rare large outliers
which correspond to large amplitude of the source. Among the wealth
of distributions that meet these requirements, we choose the
parametric family of distributions, known as the {\em generalized
gamma distribution}, $\theta_k\sim{\rm GenGamma}(r,\beta,\theta_0)$,
defined as
\begin{eqnarray}\label{generalized gamma}
\pi_{\rm hyper}(\theta) & =&\pi_{\rm hyper}(\theta;r,\beta,\theta_0)
 \nonumber\\ &\propto&  \prod_{k=1}^K \theta_k^{r\beta -1}{\rm
exp}\left(-\frac{\theta^r_k}{\theta_0^r}\right)\!=\!{\rm
exp}\left(\!-\!\sum_{k=1}^K\frac{\theta_k^r}{\theta_0^r}
+(r\beta\!-\!1)\sum_{k=1}^K\log\theta_k\right).
\end{eqnarray}
In particular, we remark that by choosing $r=1$, we have the {\em
gamma distribution}, $\theta_k\sim{\rm Gamma}(\beta,\theta_0) = {\rm
GenGamma}(1,\beta,\theta_0)$,
\[
\pi_{\rm hyper}(\theta) \propto \prod_{k=1}^K \theta_k^{\beta
-1}{\rm exp}\left(-\frac{\theta_k}{\theta_0}\right) ={\rm
exp}\left(-\sum_{k=1}^K\frac{\theta_k}{\theta_0}
+(\beta-1)\sum_{k=1}^K\log\theta_k\right),
\]
while with $r=-1$, we obtain $\theta_k\sim{\rm
InvGamma}(\beta,\theta_0)= {\rm GenGamma}(-1,\beta,\theta_0)$ which
is the {\em inverse gamma distribution}, and
\[
 \pi_{\rm hyper}(\theta)\propto \prod_{k=1}^K \theta_k^{-\beta-1}{\rm
 exp}\left(-\frac{\theta_0}{\theta_k}\right)={\rm
 exp}\left(-\sum_{k=1}^K\frac{\theta_0}{\theta_k}
 -(\beta+1)\sum_{k=1}^K\log\theta_k
 \right).
\]

For gamma and inverse gamma distributions, the parameters $\beta$
and $\theta_0$ are referred to as {\em shape parameter} and {\em
scaling parameter}, respectively. A discussion of the similarities
and differences of these two distributions, in particular with
regard to the frequency of occurrence and value of outliers, can be
found in \cite{CS_IP1}, where we also propose the following
Iterative Alternating Sequential (IAS) algorithm for computing the
MAP estimate, $(\alpha_{\rm MAP},\theta_{\rm MAP})={\rm
argmax}\big\{\pi(\alpha,\theta\mid b)\big\}$:

{\em IAS MAP estimation algorithm:}
\begin{enumerate}
\item Initialize $\theta = \theta^0$ and set $i=1$; \item Update
$\alpha$ by defining $\alpha^i ={\rm argmax}\{\pi(\alpha\mid
b,\theta^{i-1})\}$;\item Update $\theta$ by defining $\theta^i ={\rm
argmax}\{\pi(\theta\mid b,\alpha^{i})\}$; \item Increase $i$ by one
and repeat from 2. until convergence.
\end{enumerate}

In the algorithm, following Bayes' formula, the conditional
posterior probabilities are obtained as $\pi(\alpha\mid
b,\theta^{i-1})\propto \pi(\alpha,\theta^{i-1}\mid b)$ and
$\pi(\theta\mid b,\alpha^{i})\propto \pi(\alpha^i,\theta\mid b)$,
i.e., alternatingly the vector $\alpha$ or the vector $\theta$ in
the expression for the posterior density set to the current
estimate.

The IAS algorithm has been previously applied to image and signal
deblurring \cite{CS,CS_IP1,CS_IP2}, and it is found to give a fast
and stable algorithm that is easy to implement.

Before discussing how to organize the computations for different
choices of the hyperprior parameter $r$ in (\ref{generalized
gamma}), pointing out connections with known algorithms as
appropriate, we want to point out a difference between our approach
and previously proposed ones.

In the literature of Bayesian hierarchical models, the gamma
distribution is often suggested as a hyperprior for the precision,
or inverse of the variance, because of its conjugacy property. This
corresponds to the inverse gamma distribution for the variance. The
conjugacy property is useful, e.g., when variational Bayes methods,
or variants of the closely related EM algorithm, are used and
analytic marginal integrals are desired (Rao-Blackwellization), see,
e.g., \cite{liu}. Relevant references for the MEG problem are
\cite{sato,wipf}. Interestingly, in \cite{sato} the gamma hyperprior
for the precision was suggested but the connection with the
regularization methods was not pointed out. We emphasize that our
approach neither needs the conjugacy, nor does it take advantage of
it. In fact, as we will show below, the IAS algorithm yields fast,
efficient and explicit estimators with a large range of parameter
values corresponding to non-conjugate models. The generalized gamma
distribution is chosen here solely on the basis that it allows rare
outliers.

\subsection*{Gamma distribution and Minimum Current Estimate}

Consider the posterior density of the pair $(\alpha,\theta)$ when
the hyperprior is the gamma distribution
{\setlength\arraycolsep{2pt}
\begin{eqnarray*}
 \pi(\alpha,\theta\mid b)  \propto   {\rm exp} \Bigg(
 & - & \frac 1{2\sigma^2}\| b-M\alpha\|^2 \\ & - & \frac12\|D_\theta^{-1/2}
\alpha\|^2  -    \frac 1{\theta_0}\sum_{k=1}^K\theta_k +
\left(\beta-\frac 32\right)\sum_{k=1}^K\log\theta_k \Bigg).
\end{eqnarray*}
} To solve the first maximization problem in the IAS MAP estimation
algorithm, set $\theta=\theta^{i-1}$ and let the updated $\alpha^j$
be the minimizer of the negative of the log-posterior,
\begin{equation}\label{LS_min}
\alpha^i = {\rm argmin}\left(\frac 1{2\sigma^2}\| b-M\alpha\|^2 +
\frac12\|D_{\theta^{i-1}}^{-1/2} \alpha\|^2\right),
\end{equation}
which is the least squares solution of the linear system
\begin{equation}\label{LS_sol}
 \left[\begin{array}{c}\sigma^{-1} M \\ D_{\theta^{i-1}}^{-1/2}
 \end{array}\right] \alpha =
 \left[\begin{array}{c}\sigma^{-1} b \\ 0\end{array}\right].
\end{equation}
Subsequently, setting $\alpha = \alpha^{i}$, the updated value of
$\theta$ can be found by differentiating the log-posterior with
respect to $\theta$ and setting the derivative equal to zero. The
resulting equation is a second order equation
\[
 \frac 12
{\alpha_j^2}/{\theta_j^2} - 1/\theta_0
 + {\eta}/{\theta_j}
 = 0, \quad \eta = \beta-3/2, \quad \alpha_j=\alpha_j^i,
 \]
for the positive root, whose analytic expression is
\[
\theta_j^i = \frac 12\theta_0\big(\eta+\sqrt{\eta^2 +
 2\alpha_j^2/\theta_0}\big).
 \]
In particular, if we let $\eta=0$, we have $\label{update_theta}
 \theta_j^i = |\alpha_j^i| \sqrt{{\theta_0}/{2}}$,
and, by substituting this in (\ref{LS_min}), we obtain
\[
\alpha^i = {\rm argmin}\left(\frac 1{2\sigma^2}\| b-M\alpha\|^2 +
\frac1{\sqrt{2\theta_0}}\sum_{k=1}^K
\frac{\alpha_k^2}{|\alpha_k^{i-1}|}\right),
\]
which is also a fixed point iterate of the minimization problem
whose solution is the Minimum Current Estimate (MCE) \cite{uutela}
\[
 \alpha_{\rm MC}  =  {\rm argmin}\left(\| b-M\alpha\|^2 +
 \delta \sum_{k=1}^K
\frac{\alpha_k^2}{|\alpha_k|}\right) ,\quad \delta =
\sqrt{\frac{2}{\theta_0}}\sigma^2.
\]
Observe that, by choosing $\beta>3/2$ in the hyperprior, we avoid
the problem of dividing by components $\alpha_k^{i-1}$ near or equal
to zero. This hence provides a natural regularization method for
solving the Minimum Current Estimate problem.

\subsection*{Generalized gamma distribution and
$\ell^p$--estimates}

The choice of the hyperprior from the family of generalized gamma
distributions leads to the posterior model
{\setlength\arraycolsep{2pt} \begin{eqnarray*}
 \pi(\alpha,\theta\mid b)\propto{\rm exp}\Bigg( & - & \frac 1{2\sigma^2}
 \| b-M\alpha\|^2 \\ & - & \frac12\|D_\theta^{-1/2} \alpha\|^2  -  \frac
1{\theta_0^{r}}\sum_{k=1}^K\theta_k^{r} +\left(r\beta-\frac
32\right)\sum_{k=1}^K\log\theta_k\Bigg).
\end{eqnarray*}}
The updating of $\alpha$ given the current value $\theta^{i-1}$
requires solving  the system (\ref{LS_sol}) as in the case of gamma
hyperprior, while the updating formula for the variance parameter
changes. Setting the derivative of the logarithm of the posterior
density equal to zero leads to the algebraic equation
\[
\frac{1}{2}{\alpha_j^2}/{\theta_j^2} -r
{\theta_j^{r-1}}/{\theta_0^r} +({r\beta-3/2})/{\theta_j} =0, \qquad
\alpha_j=\alpha_j^i.
\]
This equation does not have, in general, a closed form solution,
although when $r\beta = 3/2$ the solution is simply
\[
\theta_j^i = ({\theta_0^r\alpha_j^2}/{2 r})^{1/(r+1)}.
\]

As in the case of the gamma distribution, after substituting this
solution into the objective function in (\ref{LS_min}) we notice
that the updated $\alpha^i$ is a fixed point iterate of the
$\ell^p$--penalized regularization problem,
\[
 \alpha_{p} ={\rm argmin}\left(\| b-M\alpha\|^2 +
\delta \sum_{k=1}^K |\alpha_k|^p\right), \quad \delta =
2\sigma^2\left(\frac{2r}{\theta_0^r}\right)^{{1}/({r+1})},\] with $r
= p/(2 -p)$. It is known that when $0<p < 1$, i.e., $0<r<1$, this
solution, like the MCE solution, tends to minimize the support of
the estimated current, thus yielding a good localization of the
focal activity. On the other hand, letting $r\to\infty$, or,
equivalently, $p\to 2$, the MAP estimate approaches the Tikhonov
regularized solution with a quadratic penalty. The intermediate case
$1<p<2$ is related to the analysis presented in \cite{auranen}.

\subsection*{Inverse gamma distribution and Minimum Support Estimate}
The posterior density for the inverse gamma hypermodel  is of the
form {\setlength\arraycolsep{2pt} \begin{eqnarray*}
 \pi(\alpha,\theta\mid b)  \propto  {\rm exp}\Bigg(& - & \frac 1{2\sigma^2}\|
 b- M \alpha\|^2 \\ & - &\frac 12\|
 D^{-1/2}_\theta \alpha\|^2  -  \theta_0\sum_{k=1}^K\frac 1{\theta_k}
  -\left(\beta+\frac 32\Bigg)\sum_{k=1}^K\log\theta_k
 \right).
 \end{eqnarray*}}
Once again, the updated value  $\alpha^i$ is found by solving
(\ref{LS_sol}) keeping $\theta$ fixed to the current value
$\theta^{i-1}$, while $\theta_j$ is the zero of the derivative of
the logarithm of the posterior which satisfies
\[
\frac 12 {\alpha_j^2}/{\theta_j^2}  + {\theta_0}/{\theta_j^2}
 -\kappa /{\theta_j}
 = 0,
 \]
with $ \kappa = \beta+3/2$, and $\alpha_j = \alpha_j^i$, and can be
expressed as
\[
\theta_j^i = \left(\frac 12 \alpha_j^2 + \theta_0\right)/\kappa.
\]
By interpreting this algorithm as a fixed point step of a
regularization scheme with a nonlinear penalty term, we can
reformulate it as the following minimization problem
\[
 \alpha  ={\rm argmin}\left(\| b-M\alpha\|^2 +
\delta \sum_{k=1}^K \frac {(\alpha_k)^2}{(\alpha_k)^2 +
2\theta_0}\right),\quad \delta = 4\kappa\sigma^2,
\]
whose solution is the Minimum Support Estimate (MSE)
\cite{nagarajan}. In \cite{CS,CS_IP1,CS_IP2}, the authors have shown
that, in the context of traditional image processing, the
corresponding penalty is related to the Perona-Malik functional
\cite{perona}.

\subsection*{Higher order inverse gamma distributions}
The gamma and inverse gamma distributions are standard models in
hierarchical Bayesian methods and the MAP estimates correspond to
known regularizing schemes. The combination of the generalized gamma
distribution and the IAS algorithm provide models that are
computationally tractable and lead to new estimators. As an example,
if we choose $r= -q$, where $q>1$ is an integer, the posterior
distribution is {\setlength\arraycolsep{2pt}
\begin{eqnarray*}
 \pi(\alpha,\theta\mid b)  \propto  {\rm exp}\Bigg(& - & \frac 1{2\sigma^2}\|
 b- M \alpha\|^2 \\ & - &\frac 12\|
 D^{-1/2}_\theta \alpha\|^2  -  \theta_0^q\sum_{k=1}^K\frac 1{\theta_k^q}
  -\left(q \beta +\frac 32\Bigg)\sum_{k=1}^K\log\theta_k
 \right).
 \end{eqnarray*}}
The algebraic equation for updating $\theta_k$ is
\[
 q\theta_0^q + \frac 12 x_k^2\theta_k^{q-1} -\left(q\beta +\frac
 32\right)\theta_k^q = 0,
\]
i.e., $\theta_k$ is a positive root of a polynomial of order $q$,
and can be computed in a stable way by considering the companion
matrix.

\section{Hyperpriorconditioners}

Direct computation of the least squares solution of the linear
system (\ref{LS_sol}), which yields the updated $\alpha$, becomes
prohibitively slow when the dimensionality of the problem is large,
as is the case in the applications that we are considering here.
Iterative methods are the method of choice for the solution of large
scale linear systems. In the case of linear discrete ill-posed
problems, this approach is particularly attractive because of the
inherent regularizing properties of iterative solvers when equipped
with suitable stopping criteria. Iterative linear system solvers
start from a given approximate solution and proceed to determine a
sequence of improved approximate solutions.

Assume, for the moment, that in the process of updating $\alpha$ we
introduce the change of variable $w=D_{\theta_{\rm c}}^{-1/2}\alpha$
where $\theta_{\rm c}$ is the current value for $\theta$, and
express the solution of the optimization step in the form
\begin{equation}\label{tikhonov}
 w_+ = {\rm argmin}\left(\frac 1{2\sigma^2}\| b-MD_{\theta_{\rm
c}}^{1/2} w\|^2 + \frac12\|w\|^2\right), \quad \alpha_+ =
D_{\theta_{\rm c}}^{1/2}w_+.
\end{equation}
We remark that while in the statistical framework a suitable choice
of the matrix $D_{\theta_{\rm c}}$ makes the change of variable
equivalent to whitening the random variable $\alpha$, in the context
of iterative linear systems solvers, this transformation amounts to
 preconditioning. Since in our problem the choice of the
matrix $D_{\theta_{\rm c}}^{1/2}$ is dictated by the selection of
the prior, following \cite{CS} we refer to it as priorconditioner,
to emphasize the connection between the numerical performance and
the statistical setting.

After the change of variables, the solution of the least squares
problem (\ref{tikhonov}) for updating $\alpha$ coincides with the
standard Tikhonov regularized solution for solving the
preconditioned linear system
\begin{equation}\label{lsqr iterative}
 \sigma^{-1}MD_{\theta_{\rm c}}^{1/2} w = \sigma^{-1}b,\quad \alpha =
D_{\theta_{\rm c}}^{1/2}w.
\end{equation}
It has been shown in the literature that iterative Krylov subspace
methods such as CGLS (see \cite{saad}) with early stopping of the
iterations may give results of comparable quality to Tikhonov
regularization  but are computationally much more efficient.

The introduction of a suitable right priorconditioner, which in the
Bayesian framework is related to the prior of the unknown of primary
interest, to bias the iterates towards a desirable subspace, has
been shown to improve the quality of the computed solution, in
particular when the number of iterations is limited by either high
computational costs or a large noise level in the data \cite{CS}.

In general, statistically inspired preconditioners, which convey
into the linear system solver prior beliefs about the solution, are
constructed from the covariance matrix of the prior. In the
application considered here, however, the prior is a function of
unknown parameters, whose distribution is, in turn, given by the
hyperprior. Since the parameters of the prior are themselves part of
the estimation problem, they are updated at each iteration step. To
ensure that the solution of the minimization problem (\ref{LS_min})
uses the most up to date information about the problem, as soon as
$\theta_{\rm c}$ becomes available, we update the priorconditioning
matrix $D^{1/2}_{\theta_{\rm c}}$, then proceed to compute a new
estimate of $\alpha$.

We remark that the idea of updating the preconditioner to take
advantage of newly acquired information about the linear systems was
already proposed in the context of a Flexible Generalized MINimal
RESidual (FGMRES) scheme \cite{saad}, although the motivation for
the updating were quite different. In the Bayesian paradigm we can
view priorconditioning in the context of hypermodels as
priorconditioning {\em conditioned on} the present estimate of the
prior parameters.

Finally, it is of interest to notice that the alternating updating
scheme where the least squares problem is solved with preconditioned
iterative methods (\ref{lsqr iterative}) includes as a special case
an algorithm previously proposed in the literature, based on
adaptive weighting of the $\ell^2$--norm. In fact, when the
hyperprior is the gamma distribution with $\beta = 3/2$, the IAS is
essentially the FOCUSS algorithm discussed, e.g. in
\cite{gorodnitsky}. While in the FOCUSS algorithm the regularization
is obtained by passing from (\ref{lsqr iterative}) to the normal
equation corresponding to (\ref{tikhonov}), here we advocate
regularization by truncated iteration. The connections between the
minimum current estimate and FOCUSS from the empirical Bayesian
point of view have been pointed out also in \cite{wipf}.

\section{MCMC and regions of interest}

A great advantage of the Bayesian approach over different
regularization schemes is that starting from the posterior density,
we can compute a number of different estimates and furthermore
quantify their reliability. The uncertainty quantification, however,
usually requires sequential sampling techniques which are
computationally considerably more intensive than optimization based
computation of single estimates, in particular when applied to a
detailed three dimensional model.

Various dimension reduction methods have been proposed in the
literature to make the MCMC sampling viable. A common approach is to
restrict the source sampling either to the surface of the brain or
to a thin cortical layer, see, e.g., \cite{sato}. In MEG, a further
reduction of the dimensionality of model may be achieved by
restricting the sampling to the cortical regions with a non-radial
normal vector, see, e.g., \cite{auranen}. These model reductions are
not applicable for us, since we consider both EEG and MEG and the
possibility of recovering deep sources with the MCMC sampling.

Fortunately, in applications where we are interested in local
sources, it is often sufficient to restrict the sampling to a much
smaller Region Of Interest (ROI), around the potentially active
area. The selection of the ROI can be based on prior information
about the expected activity: for example, the primary response to a
visual stimulus is expected to occur in the occipital lobe.
Alternatively, the ROI can be selected around an estimated focus of
activity. Note that the concept of ROI does not exclude the
possibilities of restricting the sampling to a portion of the
cortical layer, as is done in the cited articles, or of sampling
over the whole brain, if the computing resources are not an issue.

Once the ROI has been identified, we collect the indices of the
source basis vectors $\bj_k$ whose support is in the ROI in the
index vector $I_{\rm ROI}$, and the remaining ones in the vector
$I_0$. We then partition the  vectors $\alpha$ and $\theta$
accordingly, introducing the notation $\alpha_{\rm
ROI}=\alpha(I_{\rm ROI})$, $\alpha_0=\alpha(I_0)$, $\theta_{\rm
ROI}=\theta(I_{\rm ROI})$ and $\theta_0=\theta(I_0)$.

We can now perform MCMC sampling over the ROI, fixing the outside
current values and prior variance to prescribed values $\alpha_0$,
$\theta_0$, using the conditional distribution,
\[
 \pi(\alpha_{\rm ROI},\theta_{\rm ROI}\mid \alpha_0,\theta_0,b)
 \propto \pi([\alpha_{\rm ROI},\alpha_0],[\theta_{\rm ROI},\theta_0]\mid
 b).
\]
Evidently, $\alpha_0=0$ is the most natural choice, corresponding to
the assumption that no activity outside the ROI appears. The MCMC
algorithm that we propose is an independence sampling method, where
$\alpha_{\rm ROI}$ and $\theta_{\rm ROI}$ are updated sequentially
by a procedure analogous to that of the IAS estimation algorithm.
The updating of $\alpha_{\rm ROI}$ is done using the conditional
normality of the posterior, while the updating of $\theta_{\rm ROI}$
is done via a full scan Gibbs sampler \cite{CS,KS,liu}.

{\em MCMC sampling over ROI:}
\begin{enumerate}
\item Initialize $\alpha_{\rm ROI}^0$, $\theta_{\rm ROI}^0 $ and
set $i=1$. Define $M$, the desired sample size. \item Draw
$\alpha_{\rm ROI}^i$ from the Gaussian density
\[
 \pi(\alpha_{\rm ROI}\mid\theta_{\rm
 ROI}^{i-1},\alpha_0,\theta_0,b)\propto
 \pi([\alpha_{\rm ROI},\alpha_0],[\theta_{\rm ROI}^{i-1},\theta_0]\mid
 b).
\]
\item Draw $\theta_{\rm ROI}^i$ componentwise with a Gibbs sampler
from the density
\[
 \pi(\theta_{\rm ROI}\mid\alpha_{\rm
 ROI}^{i},\alpha_0,\theta_0,b)\propto
 \pi([\alpha_{\rm ROI}^i,\alpha_0],[\theta_{\rm ROI},\theta_0]\mid
 b).
\]
\item If $i=M$, stop; otherwise increase $i$ by one and repeat
from 2.
\end{enumerate}

In the practical implementation of the algorithm we update
$\alpha_{\rm ROI}$ by defining a matrix $G$ and its partitioning,
\[
 G = \left[\begin{array}{c}\sigma^{-1} M \\
D_{\theta^{i-1}}^{-1/2}
 \end{array}\right] = \left[\begin{array}{cc} G_{\rm ROI} &
 G_0\end{array}\right],
\]
where $G_{\rm ROI}$ contains the columns of $G$ with indices in
$I_{\rm ROI}$ and $G_0$ the remaining columns. The updated
$\alpha_{\rm ROI}^i$ is obtained by solving, in the least squares
sense, the linear system
\[
 G_{\rm ROI}\alpha_{\rm ROI} = \left[\begin{array}{c} \sigma^{-1}
 b \\ 0 \end{array}\right] - G_0\alpha_0 + w,\quad w\sim{\mathcal
 N}(0,I).
\]
The updating of $\theta_k$, $k\in I_{\rm ROI}$, is performed by
drawing from the one-di\-men\-sio\-nal probability density
\[
\pi_k(\theta_k)\propto {\rm exp}\left( -
\frac{\alpha^2_k}{2\theta_k}
-\left(\frac{\theta_k}{\theta_0}\right)^r +\left(r\beta-\frac
32\right)\log\theta_k\right)
\]
by the inverse cumulative distribution method \cite{CS,KS}. Hence,
the sampling technique just outlined takes advantage of the
conditional normality of the prior and of the mutual independence of
the variances, similarly to the IAS algorithm.

\section{Computed experiments}

In this section we apply the methodology derived above to inverse
source problems by first considering an example with a simplified
planar geometry using a traditional singular dipole model, then
applying it to a realistic conductivity model for the human head. In
the latter case, both the MEG and EEG modalities are considered.
Since the finite element method (FEM) is needed to solve the
potential distribution, we use FEM basis functions also for
representation of the current density.

\subsection*{MEG in planar geometry}

\begin{figure}[t]
\centerline{\includegraphics[width=12.9cm]{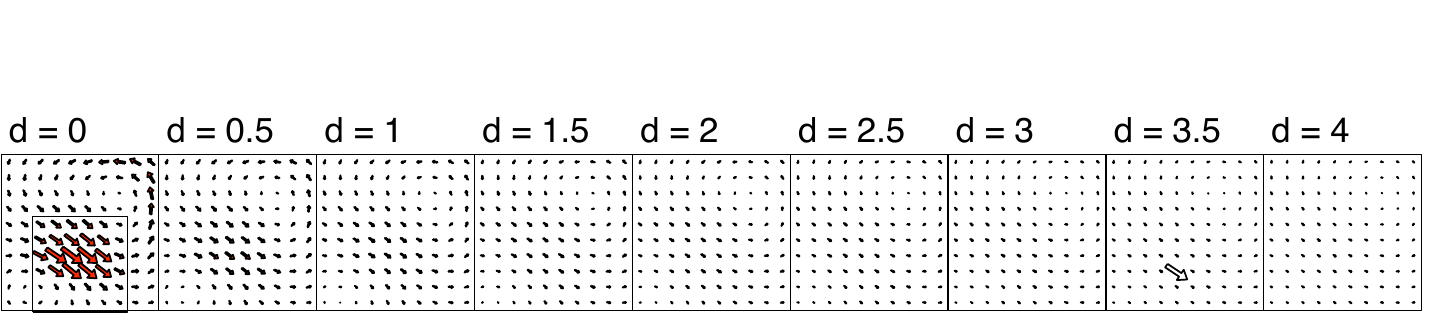}}
\centerline{\includegraphics[width=12.9cm]{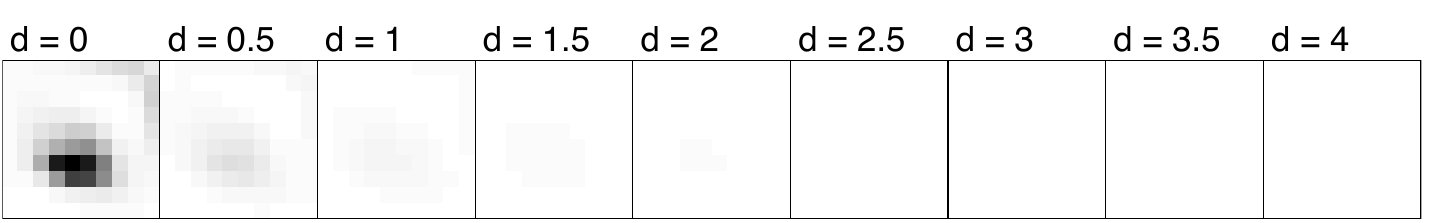}}
\caption{\label{fig:MAP layered gamma} MAP estimates of the current
(top row) and of the variance of the prior (bottom row) at different
depth in the ROI. The hyperprior is the gamma distribution. The true
current dipole used for generating simulated data is shown as a
hallow arrow, and the ROI is marked on the superficial layer.}
\end{figure}
\begin{figure}[t]
\centerline{\includegraphics[width=12.9cm]{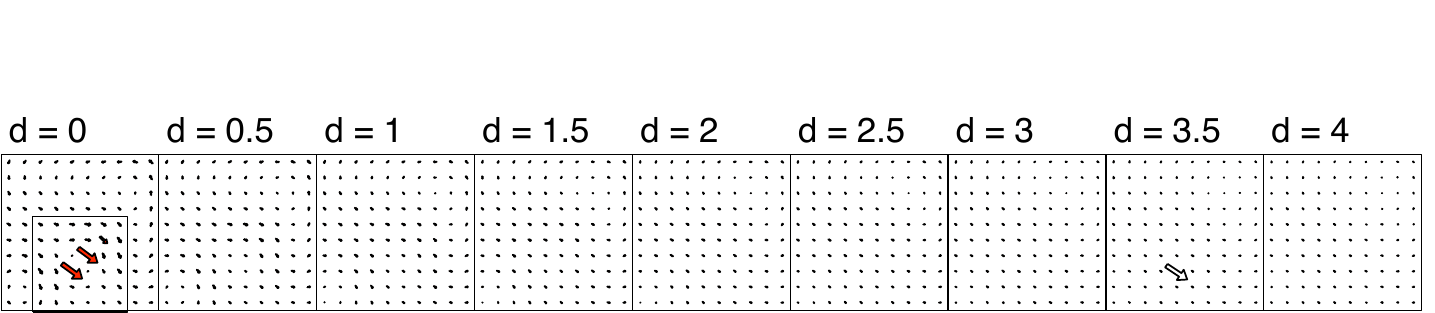}}
\centerline{\includegraphics[width=12.9cm]{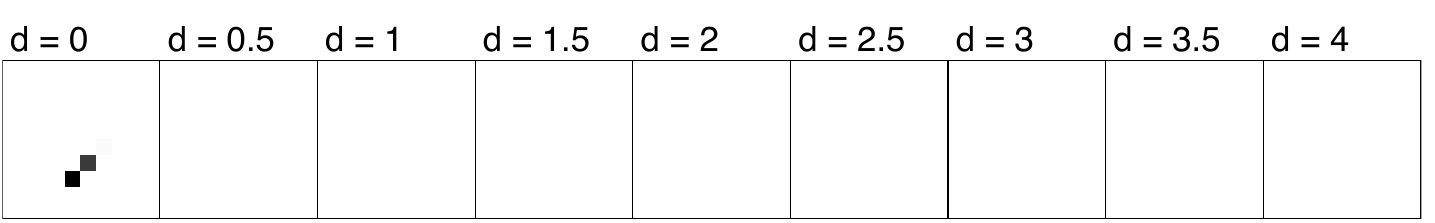}}
\caption{\label{fig:MAP layered invgamma} MAP estimates of the
current (top row) and of the variance of the prior (bottom row) at
different depth in the ROI. The hyperprior is the inverse gamma
distribution. The true current dipole used for generating simulated
data is shown as a hallow arrow, and the ROI is marked on the
superficial layer.}
\end{figure}
\begin{figure}[t]
\centerline{\includegraphics[width=12.9cm]{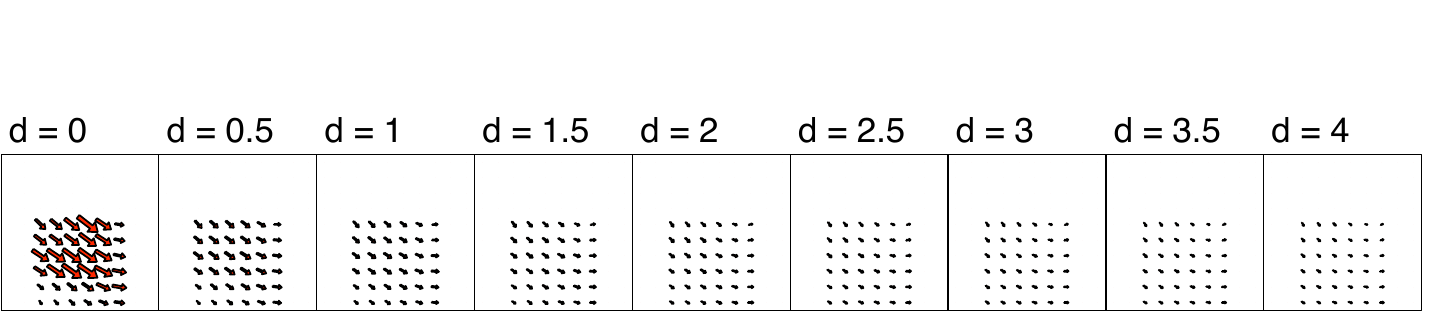}}
\centerline{\includegraphics[width=12.9cm]{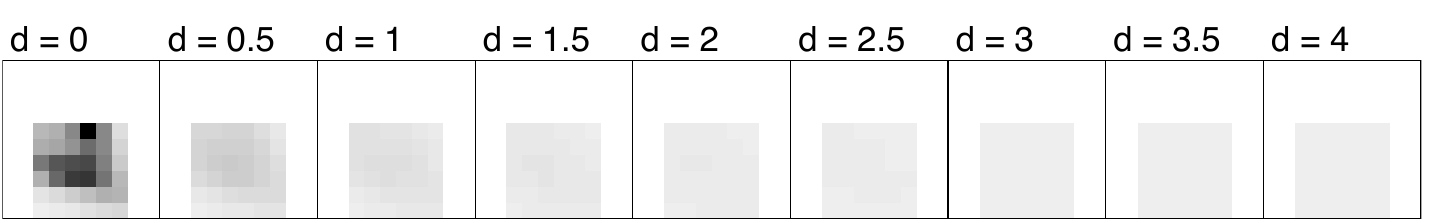}}
\centerline{\includegraphics[width=12.9cm]{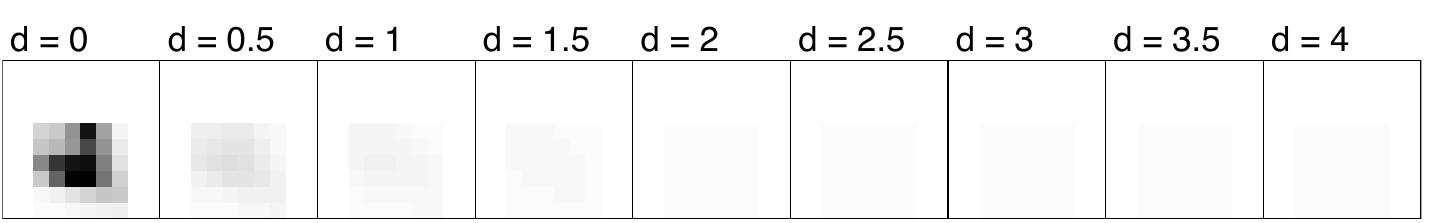}}
\caption{\label{fig:layered gamma} Posterior mean estimates of the
current (top row), of the variance of the prior (center row), and of
the variance of the norm of the current estimated over the sample
(bottom row). The hyperprior is the gamma distribution.}
\end{figure}
\begin{figure}[t]
\centerline{\includegraphics[width=12.9cm]{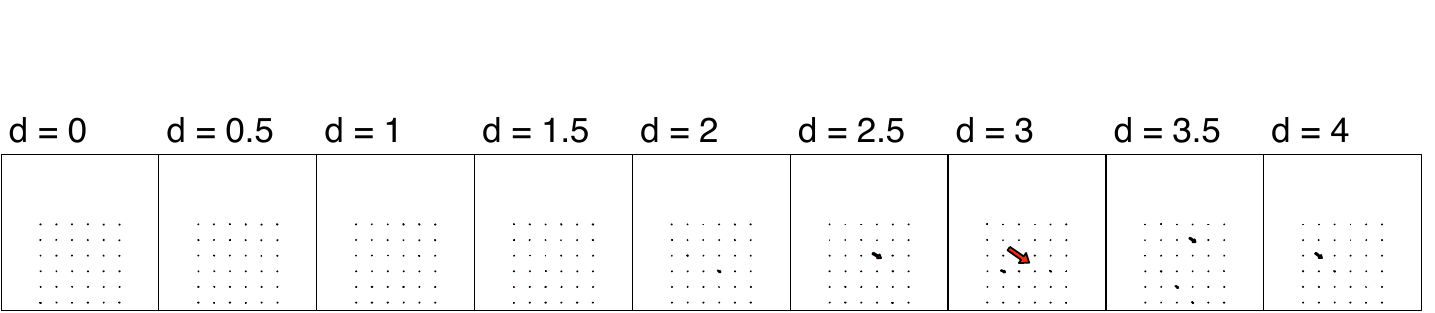}}
\centerline{\includegraphics[width=12.9cm]{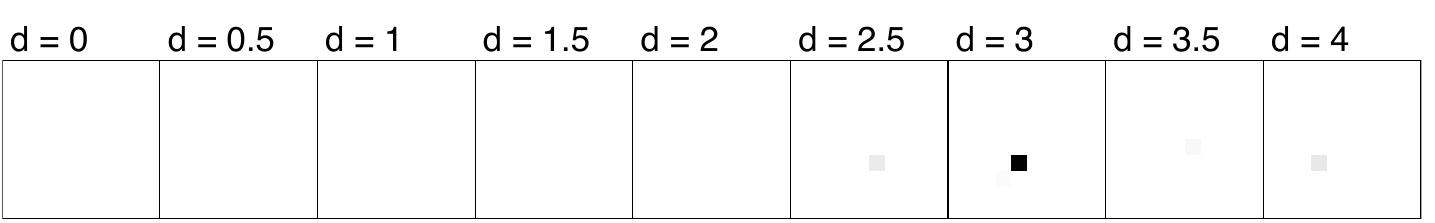}}
\centerline{\includegraphics[width=12.9cm]{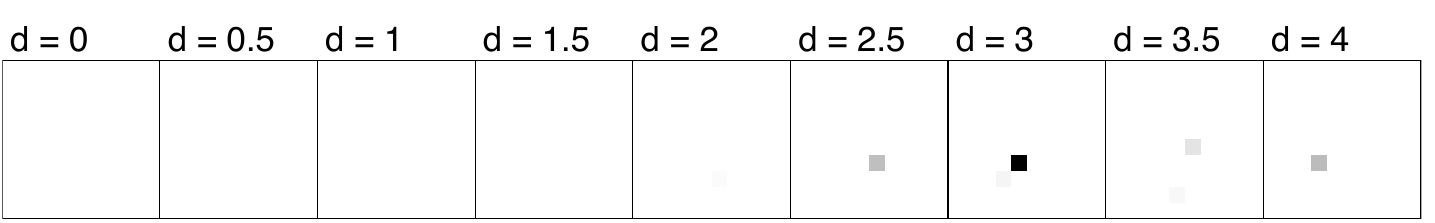}}
\caption{\label{fig:layered invgamma} Posterior mean estimates of
the current (top row), of the variance of the prior (center row) and
the variance of the norm of the current estimated over the sample
(bottom row). The hyperprior is the inverse gamma distribution.}
\end{figure}
\begin{figure}[t]
\centerline{\includegraphics{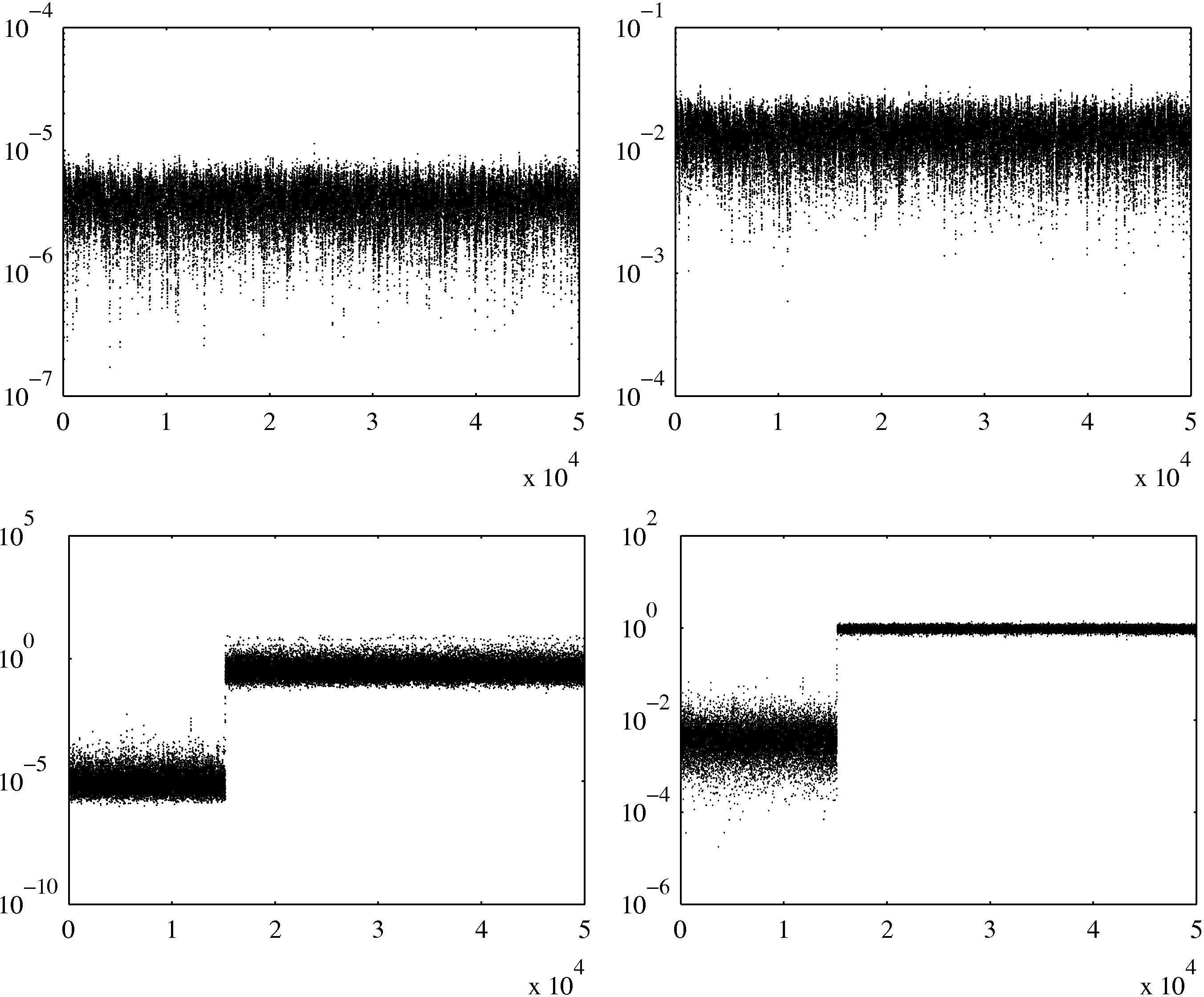}}
\caption{\label{fig:sample history} Sample histories of the
components $\theta_j$ and $\|{\bf q}_j\|$ for the gamma hyperprior
(top row) and the inverse gamma (bottom row), where $j$ is the index
of the component of maximum value of $\theta_{CM}$.}
\end{figure}

Consider a half space as a local model for the human head. The half
space model is particularly appropriate to illustrate the depth
resolution with different hypermodel parameters and with different
statistical estimators.

The magnetic field component perpendicular to the surface is
recorded in a rectangular array of observation points $x_\ell$ above
the surface $z=0$. We represent the current density as a linear
combination of point-wise current dipoles,
\[
 \J(y) = \sum_{j=1}^K \delta(y-y_k)(\alpha_k^1\be_1
 +\alpha_k^2\be_2) =\sum_{j=1}^K\delta(y-y_k)\bq_k,
\]
where $\be_1$ and $\be_2$ are orthogonal basis vectors parallel to
the plane $z=0$. In this example we ignore the volume currents,
which is tantamount to setting $\sigma=0$. More generally, assuming
a conductivity density that depends only on the depth \cite{sarvas},
we arrive at a particularly simple magnetic lead field model,
\[
 b_\ell = \frac{\mu_0}{4\pi}\sum_{k=1}^K\sum_{j=1}^2
 \frac{\be_3\cdot(\be_j\times(x_\ell -
 y_k))}{|x_\ell -y_k|^3}\alpha_k^j,
\quad \hbox{or} \quad
 b = \left[\begin{array}{cc} M^{\rm m}_1 & M^{\rm
 m}_2\end{array}\right]\left[\begin{array}{c}\alpha^1 \\
 \alpha^2\end{array}\right],\]
with $\alpha = [\alpha^1; \alpha^2] \in \R^{2K}$. When writing the
conditional prior, we identify the variances $\theta_k^1$ and
$\theta_k^2$ of the two mutually perpendicular dipoles whose
locations coincide, and the unknown variance
$\theta_k^1=\theta_k^2=\theta_k$ is a vector of length $K$. The
posterior density then becomes {\setlength\arraycolsep{2pt}
\begin{eqnarray*}
 \pi(\alpha,\theta\mid b)
\propto{\rm exp}\Bigg(& -&\frac 1{2\sigma^2}\|
 b-M\alpha\|^2  \\ & - & \frac 12
\sum_{j=1}^K \frac{\|\bq_k\|^2}{\theta_k}
 -\sum_{k=1}^K\left(\frac {\theta_k}{\theta_0}\right)^r
 +\left(r\beta - \frac{3}{2} \right)\sum_{k=1}^K\log\theta_k
 \Bigg),
 \end{eqnarray*}}
with $\|\bq_k\|^2 = (\alpha_k^1)^2 + (\alpha_k^2)^2$. We use the IAS
algorithm to calculate the MAP estimate, and  MCMC sampling over the
ROI to estimate the CM and obtain a measure of its uncertainty.

The geometry of our model consists of a rectangular array of
$10\times 10$ vertical magnetometers 2 cm above the half space, with
a distance between adjacent magnetometers of 1 cm. The dipoles are
located below the magnetometers in nine horizontal layers, each
containing a $10\times 10$ rectangular array of dipoles. The depth
of these layers varies from zero (superficial sources) to 4 cm, with
a distance between the layers of 0.5 cm.

Since the MAP estimation algorithm is tantamount to fixed point
iteration with localizing penalties, we expect good performance at
detecting focal sources of known depth. It is well known
\cite{hamalainen,lin,uutela} that when the depth of the source is
unknown, the minimum current and minimum norm estimates due to their
tendency to bias towards superficial sources may lead to gross
misplacements of the deep focal sources. We expect the same behavior
from the MAP estimates of our hypermodel. We test this by generating
synthetic data in which a single dipole source is placed 3.5 cm
below the surface of the half space. The standard deviation in the
likelihood model was assumed to be 5\% of the maximum noiseless
signal. In this simulation, we did not add artificial noise to the
data, since we are only interested in the model bias, not in the
noise sensitivity.

The MAP estimates for the dipole fields as well as for the prior
variance $\theta$ with model parameter values $r=1$ (gamma) and
$r=-1$ (inverse gamma) are shown in Figures~\ref{fig:MAP layered
gamma} -- \ref{fig:MAP layered invgamma}. When $r=1$, we use the
values $\theta_0=10^{-7}$ and $\beta = 3$ for the scaling parameter
and the shape parameter, while when $r=-1$, we set
$\theta_0=10^{-5}$ and $\beta = 3$. In both cases we perform 15
iterations with the IAS algorithm.

As expected, both hypermodels favor superficial sources, with a
relatively good localization in the horizontal direction, i.e., the
MAP estimate of the activity is above the true source. The major
differences between the two hypermodels are in the convergence rate
and in the focality of the MAP estimate. With the inverse gamma
hypermodel the iterative algorithm converges faster than with the
gamma hypermodel, in particular for non-superficial sources, and
seeks to explain the data with fewer active superficial dipoles.

To reduce the dimensionality of the sampling space, we select the
ROI to be a cylinder with a $6\times 6$ cm$^2$ square base around
the estimated superficial focal activity, shown in
Figure~\ref{fig:MAP layered gamma} -- \ref{fig:MAP layered
invgamma}, containing $9\times 6\times 6=324$ dipoles. For each
hyperparameter model we generate a sample of size $M=50\,000$,
conditional on the currents vanishing outside the ROI, and calculate
estimates of the posterior means of the vectors $\alpha$ and
$\theta$, $\alpha_{\rm CM}^j = \frac 1M\sum_{i=1}^M \alpha^{j,i}$,
$j=1,2$, and $\theta_{\rm CM} = \frac 1M\sum_{i=1}^M \theta^i$, and
of the posterior variance of the dipole amplitudes,  ${\rm
Var}(\|\bq_k\|) = \frac 1M\sum_{i=1}^M\big\{(\alpha^{1,i} -
 \alpha_{\rm CM}^1)^2 +(\alpha^{2,i} -
 \alpha_{\rm CM}^2)^2\big\}$.

Figure~\ref{fig:layered gamma} displays the plots of the posterior
mean of the current and the estimates of the prior variance, and the
posterior variance of the amplitude with the gamma hyperprior, while
Figure~\ref{fig:layered invgamma} shows the analogous results for
the inverse gamma hyperprior. The results with the two hyperpriors
are qualitatively very different. The posterior mean of the current
for the gamma hyperprior model is biased towards the surface and is
very similar to the MAP estimate, while the inverse gamma hypermodel
has a good depth resolution, with an error between the true and the
estimated source depth of 0.5 cm. The observation that the gamma
hyperprior leads to a posterior density that is qualitatively closer
to the Gaussian prior, whose mean and maximum coincide, is in line
with the fact that, as $r\to\infty$, the MAP estimate approaches the
minimum norm, or Tikhonov regularized, solution. The latter
corresponds to an $\ell^2$ prior model.

The sample histories of single components $\|{\bf q}_j\| =
((\alpha_j^1)^2 + (\alpha_j^2)^2)^{1/2}$ and $\theta_j$ reveal an
interesting feature of the posterior density. Figure~\ref{fig:sample
history} shows the sample histories of the components corresponding
to the maximum value of $\theta_{\rm CM}$. The sample histories with
the gamma hyperprior exhibit good mixing, while those corresponding
to the inverse gamma distribution seem to suggest a bimodal
posterior density.

Note that the posterior mean of the prior variance $\theta$ is in
good agreement with the posterior variance of the current amplitude,
implying that the reliability of the posterior mean current in this
geometry could be assessed directly from the mean of the variance
parameter $\theta$.

\subsection*{EEG/MEG in realistic geometry}

\begin{table}
\caption{\label{table: conductivities}The domains of the head model
and respective conductivities.}
\begin{center}
\begin{tabular}{|c|c|}
\hline
Layer & Conductivity\\
\hline
scalp &  0.33\\
\hline
skull & 0.0042\\
\hline
cerebrospinal fluid & 1 \\
\hline
brain & 0.33 \\
\hline
\end{tabular}
\end{center}
\end{table}
\begin{figure}[t]
\centerline{\begin{minipage}{4cm}
\centerline{{\includegraphics[width=3.7cm]{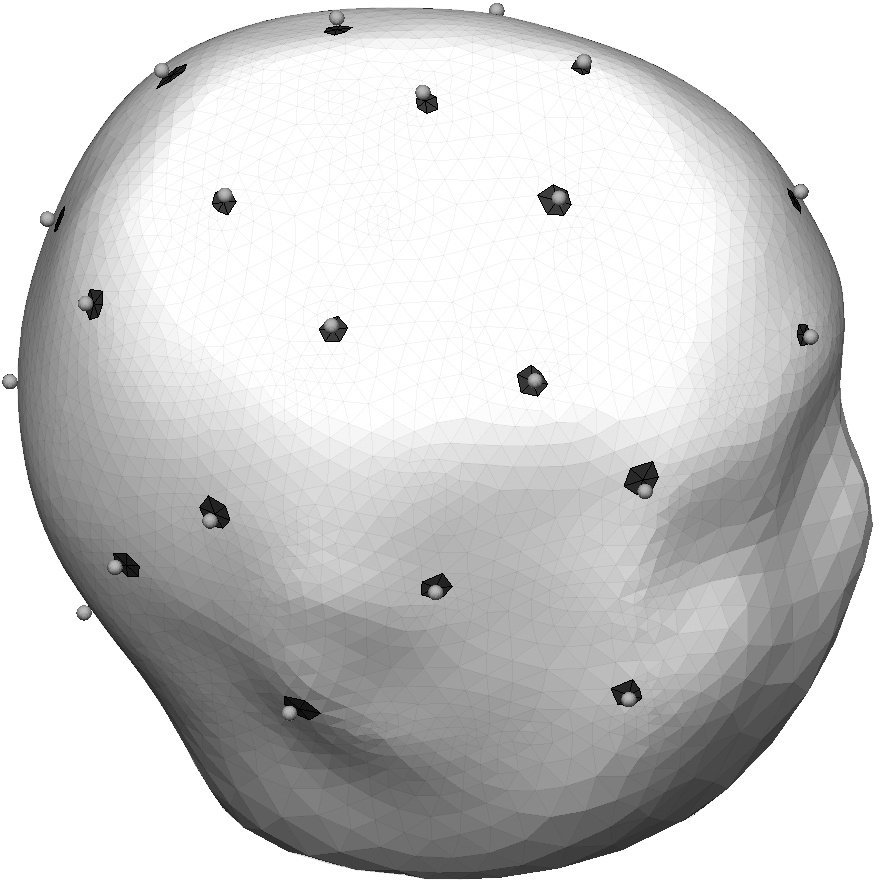}}}
\end{minipage}
\begin{minipage}{4cm}
\centerline{{\includegraphics[width=2.9cm]{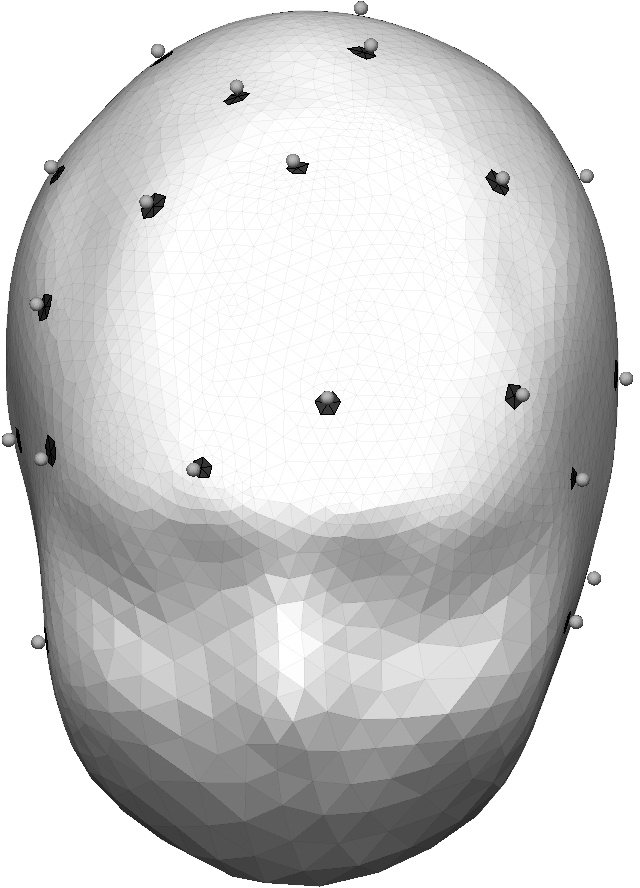}}}
\end{minipage}
\begin{minipage}{4cm}
\centerline{{\includegraphics[width=3.7cm]{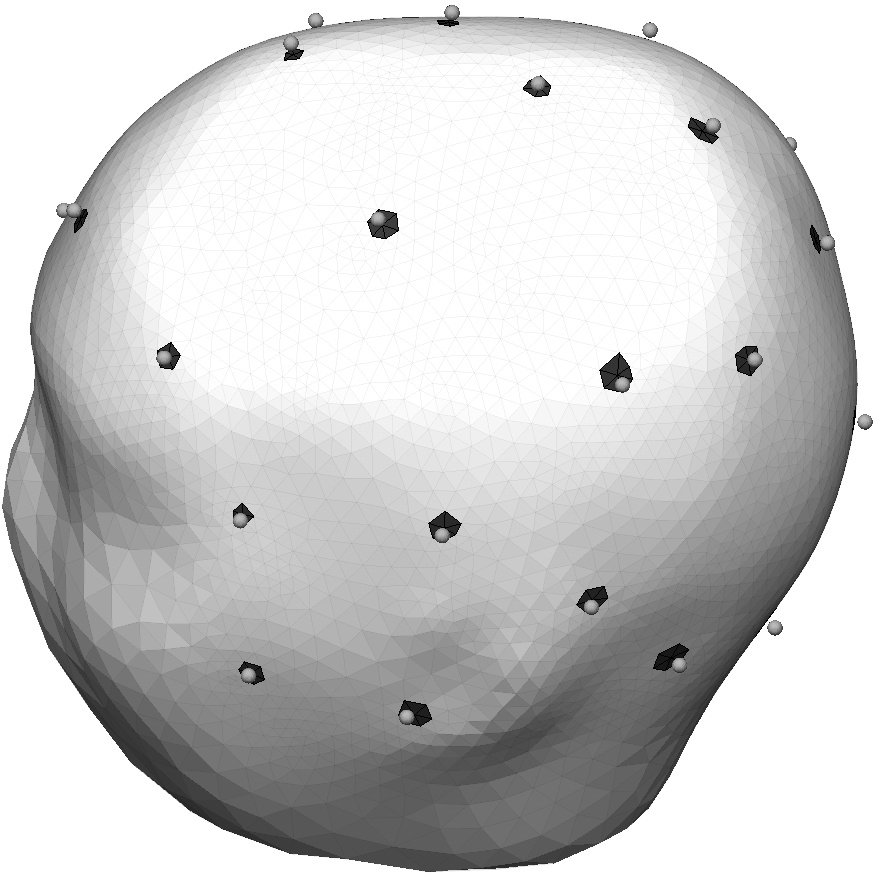}}}
\end{minipage}}
\centerline{\begin{minipage}{4cm}
\centerline{{\includegraphics[width=2.7cm]{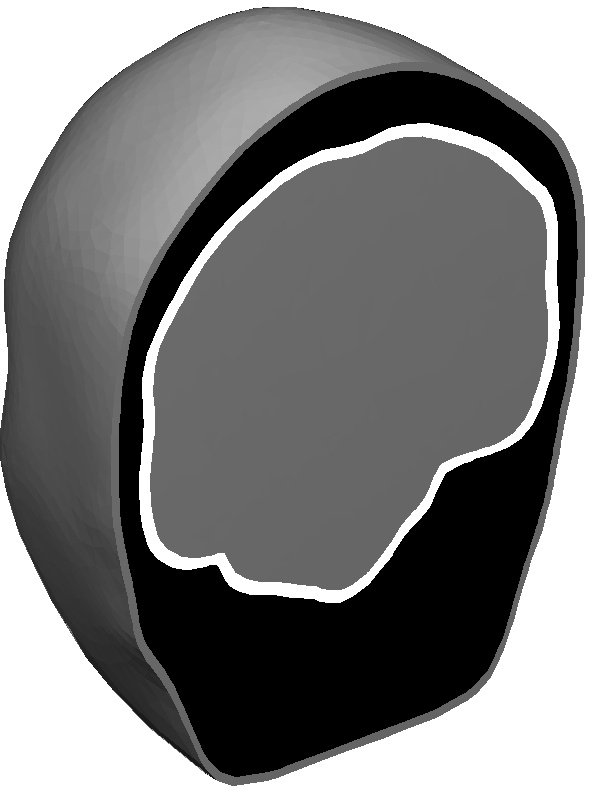}}}
\end{minipage}
\begin{minipage}{4cm}
\centerline{{\includegraphics[width=2.7cm]{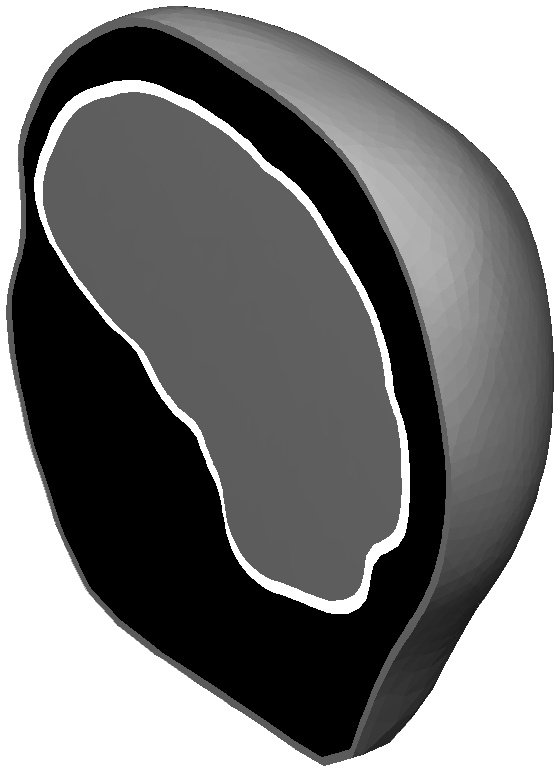}}}
\end{minipage}
\begin{minipage}{4cm}
\centerline{{\includegraphics[width=2.9cm]{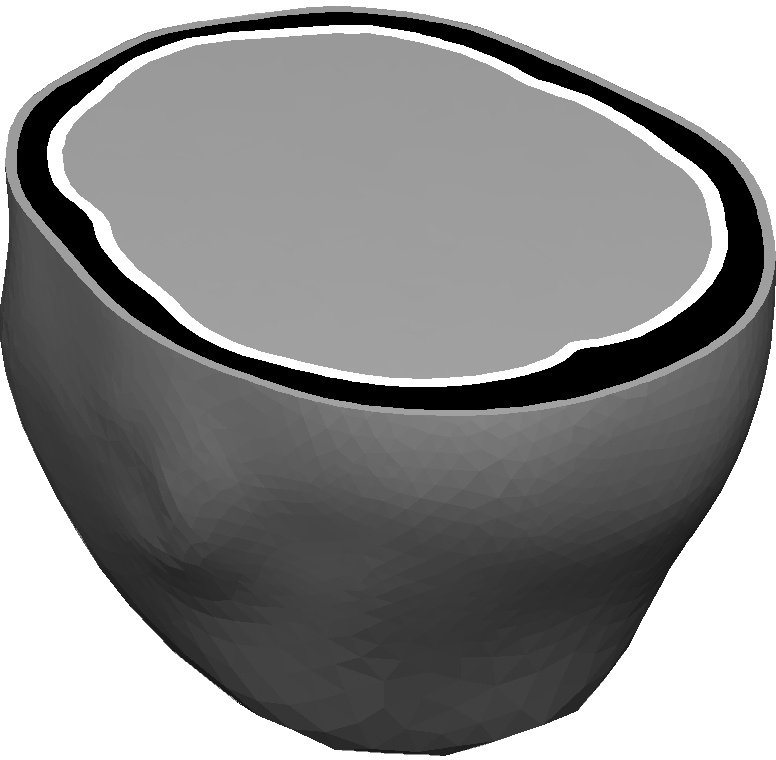}}}
\end{minipage}}
\caption{\label{fig:head geometry} Sagittal, coronal, and axial
projections (left to right) of the head model (top row) and the
conductivity distribution inside it (bottom row). The electric field
is measured using 31 contact electrodes marked by dark grey surface
patches. The 31 magnetic field measurement locations are indicated
by the lighter dark spheres over the patches.}
\end{figure}
\begin{figure}[t]
\centerline{\begin{minipage}{4cm}
\centerline{{\includegraphics[width=3.7cm]{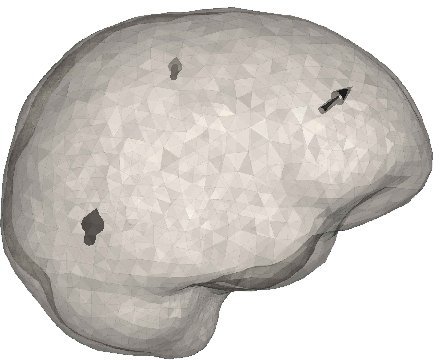}}}
\end{minipage}
\begin{minipage}{4cm}
\centerline{{\includegraphics[width=3.1cm]{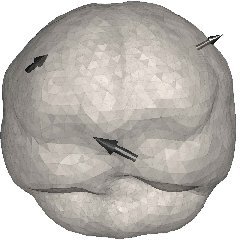}}}
\end{minipage}
\begin{minipage}{4cm}
\centerline{{\includegraphics[width=3.7cm]{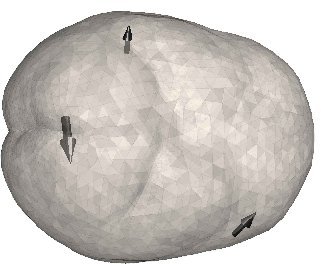}}}
\end{minipage}}
\centerline{\begin{minipage}{4cm}
\centerline{{\includegraphics[width=3.7cm]{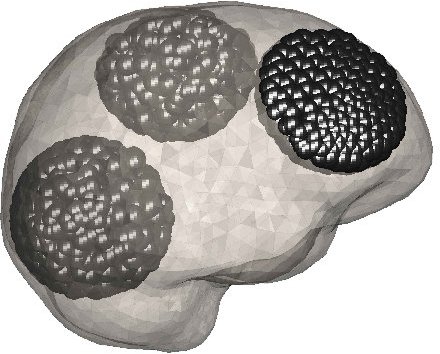}}}
\end{minipage}
\begin{minipage}{4cm}
\centerline{{\includegraphics[width=3.1cm]{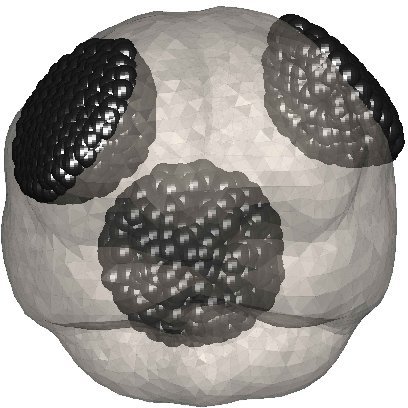}}}
\end{minipage}
\begin{minipage}{4cm}
\centerline{{\includegraphics[width=3.7cm]{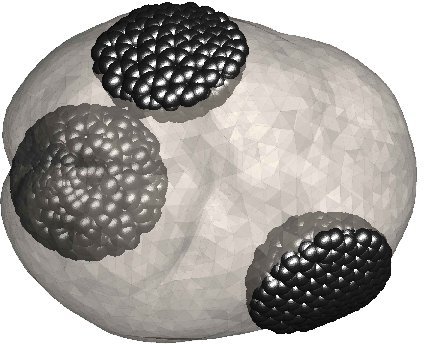}}}
\end{minipage}}
\caption{\label{fig: reference} Sagittal, coronal, and axial
projections (left to right) of the reference current density (top
row) and of the region of interest (bottom row). }
\end{figure}
\begin{figure}[t]
\centerline{\begin{minipage}{4cm}
\centerline{{\includegraphics[width=3.7cm]{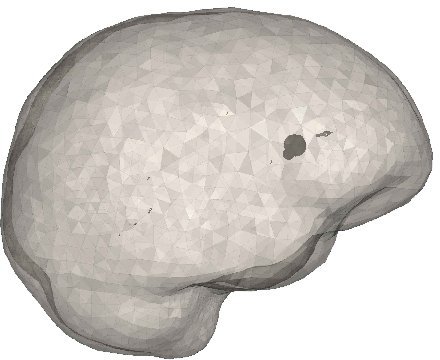}}}
\end{minipage}
\begin{minipage}{4cm}
\centerline{{\includegraphics[width=3.1cm]{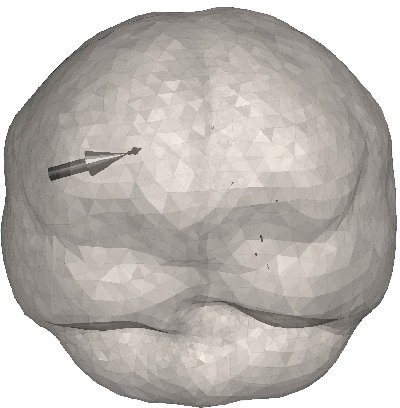}}}
\end{minipage}
\begin{minipage}{4cm}
\centerline{{\includegraphics[width=3.7cm]{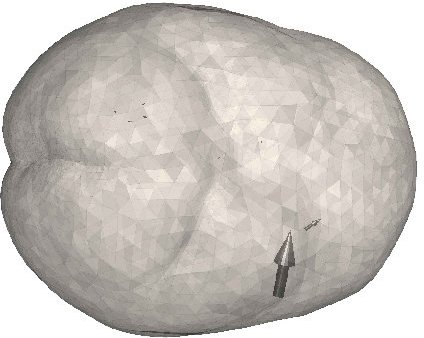}}}
\end{minipage}}
\centerline{\begin{minipage}{4cm}
\centerline{{\includegraphics[width=3.7cm]{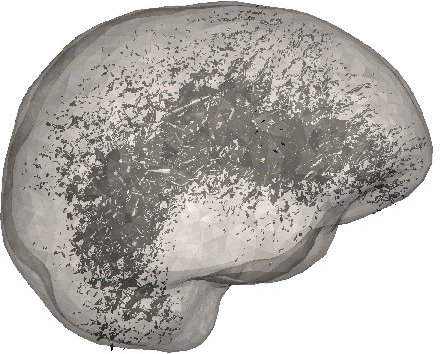}}}
\end{minipage}
\begin{minipage}{4cm}
\centerline{{\includegraphics[width=3.1cm]{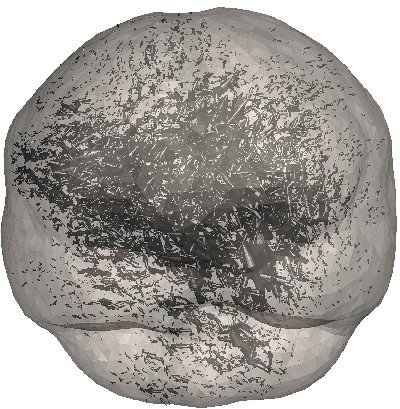}}}
\end{minipage}
\begin{minipage}{4cm}
\centerline{{\includegraphics[width=3.7cm]{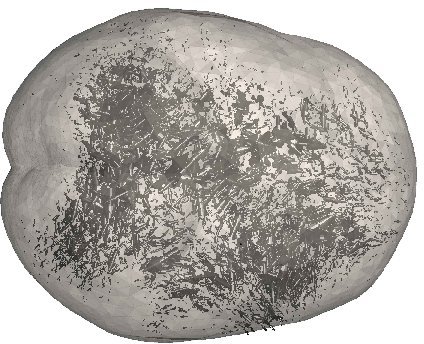}}}
\end{minipage}}
\caption{\label{fig:estimates alpha1}Sagittal, coronal, and axial
projections (left to right) of the MAP estimate from EEG data of the
current for the gamma hypermodel (top row) and of the inverse gamma
hypermodel (bottom row). To improve the readability of  the three
dimensional plots, only the current elements whose amplitude is
above 5\% of the maximum of the amplitudes in the estimate were
plotted.}
\end{figure}
\begin{figure}[t]
\centerline{\begin{minipage}{4cm}
\centerline{{\includegraphics[width=3.7cm]{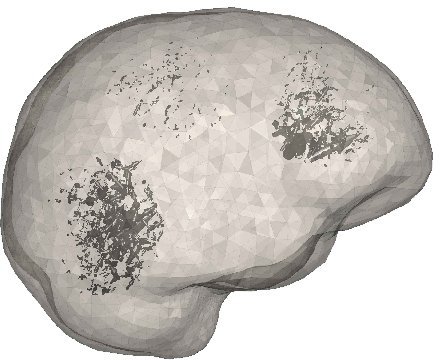}}}
\end{minipage}
\begin{minipage}{4cm}
\centerline{{\includegraphics[width=3.1cm]{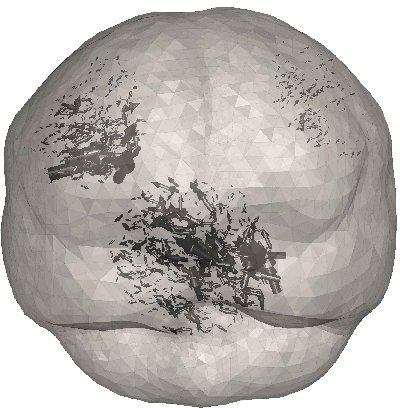}}}
\end{minipage}
\begin{minipage}{4cm}
\centerline{{\includegraphics[width=3.7cm]{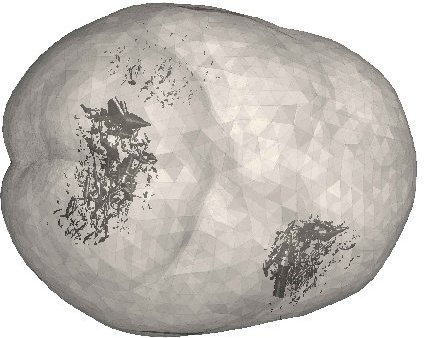}}}
\end{minipage}}
\centerline{\begin{minipage}{4cm}
\centerline{{\includegraphics[width=3.7cm]{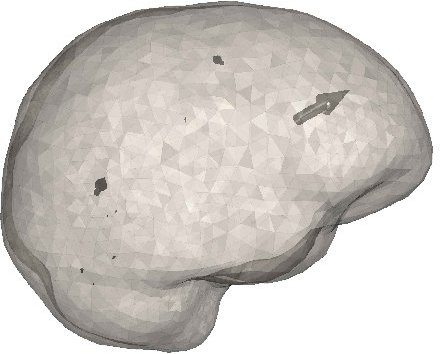}}}
\end{minipage}
\begin{minipage}{4cm}
\centerline{{\includegraphics[width=3.1cm]{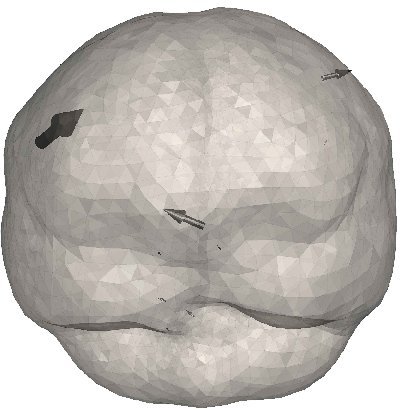}}}
\end{minipage}
\begin{minipage}{4cm}
\centerline{{\includegraphics[width=3.7cm]{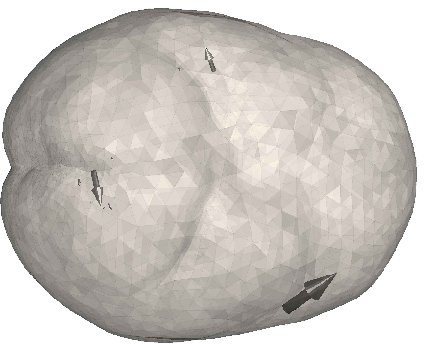}}}
\end{minipage}}
\caption{\label{fig:estimates alpha2}Sagittal, coronal, and axial
projections (left to right) of the CM estimate from EEG data of the
current for the gamma hypermodel (top row) and for the inverse gamma
hypermodel (bottom row).}
\end{figure}
\begin{figure}[t]
\centerline{\begin{minipage}{4cm}
\centerline{{\includegraphics[width=3.7cm]{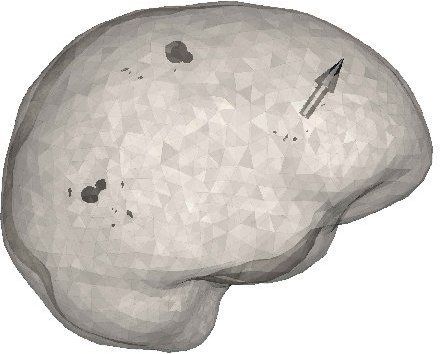}}}
\end{minipage}
\begin{minipage}{4cm}
\centerline{{\includegraphics[width=3.1cm]{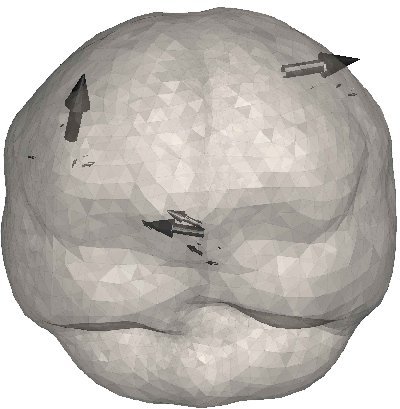}}}
\end{minipage}
\begin{minipage}{4cm}
\centerline{{\includegraphics[width=3.7cm]{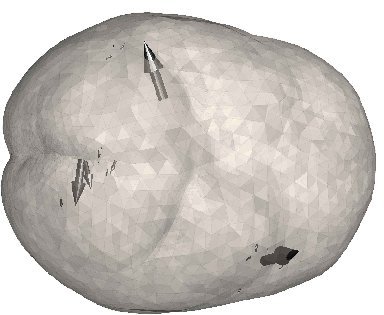}}}
\end{minipage}}
\centerline{\begin{minipage}{4cm}
\centerline{{\includegraphics[width=3.7cm]{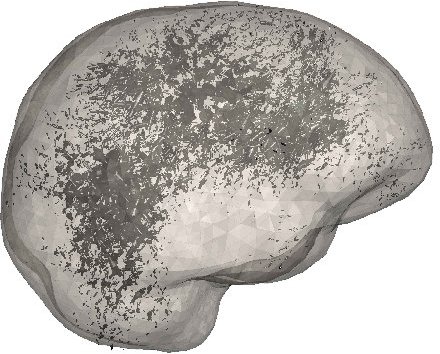}}}
\end{minipage}
\begin{minipage}{4cm}
\centerline{{\includegraphics[width=3.1cm]{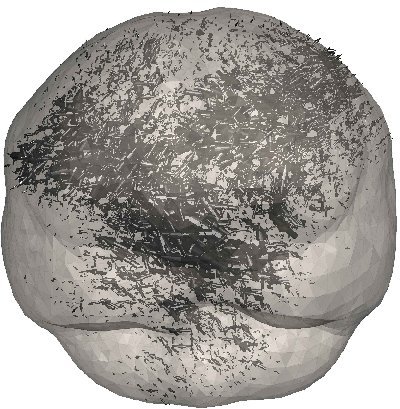}}}
\end{minipage}
\begin{minipage}{4cm}
\centerline{{\includegraphics[width=3.7cm]{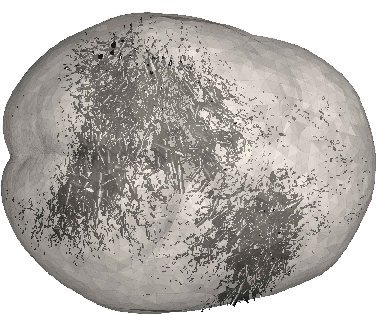}}}
\end{minipage}}
\caption{\label{fig:estimates alpha3}Sagittal, coronal, and axial
projections (left to right) of the MAP estimate from MEG data of the
current for the gamma hypermodel (top row) and for the inverse gamma
hypermodel (bottom row).}
\end{figure}
\begin{figure}[t]
\centerline{\begin{minipage}{4cm}
\centerline{{\includegraphics[width=3.7cm]{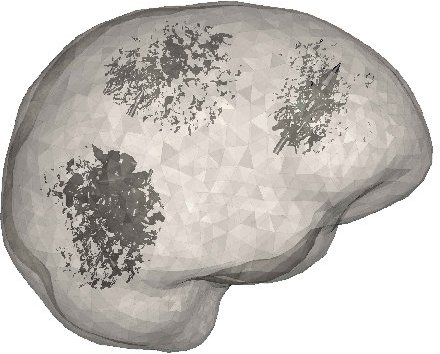}}}
\end{minipage}
\begin{minipage}{4cm}
\centerline{{\includegraphics[width=3.1cm]{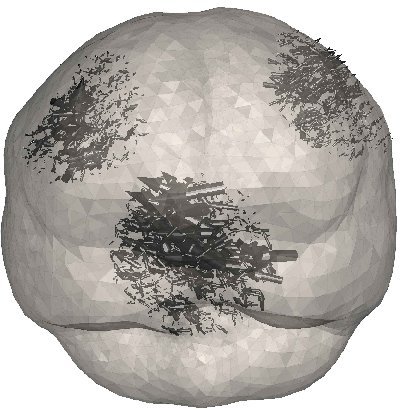}}}
\end{minipage}
\begin{minipage}{4cm}
\centerline{{\includegraphics[width=3.7cm]{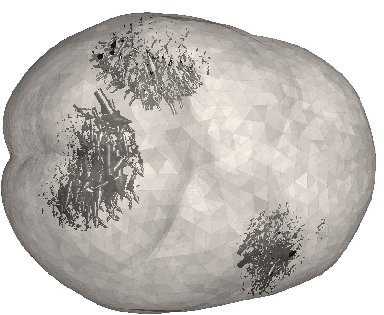}}}
\end{minipage}}
\centerline{\begin{minipage}{4cm}
\centerline{{\includegraphics[width=3.7cm]{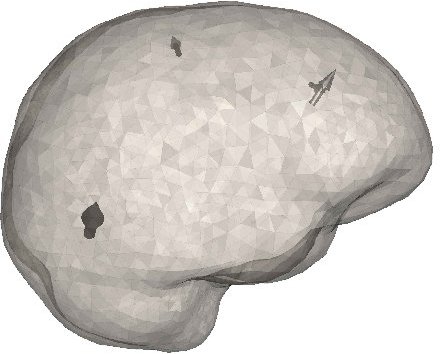}}}
\end{minipage}
\begin{minipage}{4cm}
\centerline{{\includegraphics[width=3.1cm]{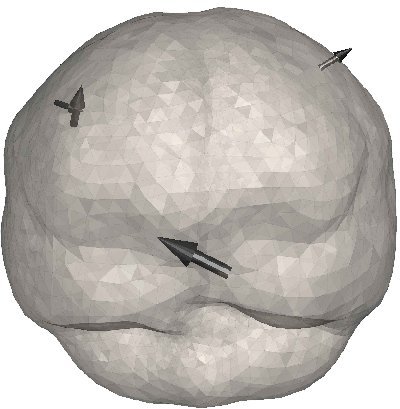}}}
\end{minipage}
\begin{minipage}{4cm}
\centerline{{\includegraphics[width=3.7cm]{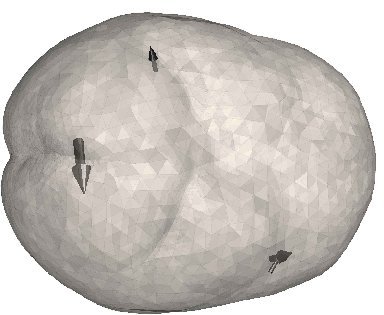}}}
\end{minipage}}
\caption{\label{fig:estimates alpha4}Sagittal, coronal, and axial
projections (left to right) of the CM estimate from MEG data of the
current for the gamma hypermodel (top row) and for the inverse gamma
hypermodel (bottom row).}
\end{figure}
\begin{figure}[t]
\centerline{\begin{minipage}{4cm}
\centerline{{\includegraphics[width=3.7cm]{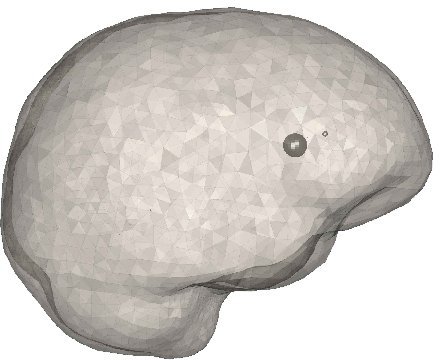}}}
\end{minipage}
\begin{minipage}{4cm}
\centerline{{\includegraphics[width=3.1cm]{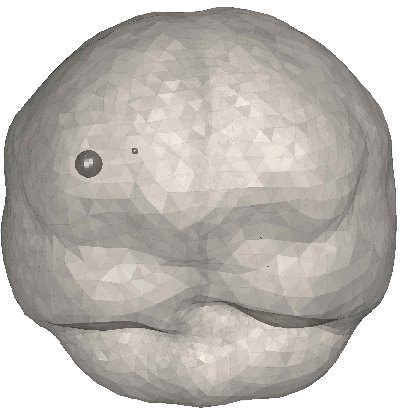}}}
\end{minipage}
\begin{minipage}{4cm}
\centerline{{\includegraphics[width=3.7cm]{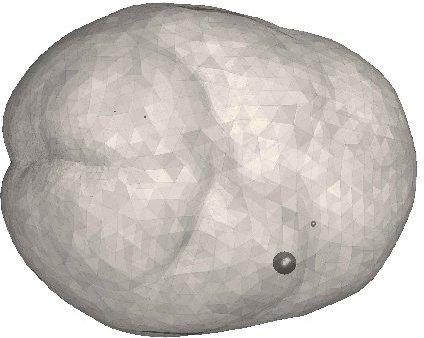}}}
\end{minipage}}
\centerline{\begin{minipage}{4cm}
\centerline{{\includegraphics[width=3.7cm]{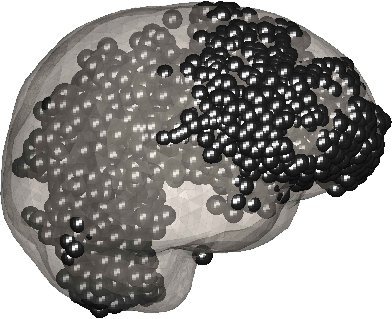}}}
\end{minipage}
\begin{minipage}{4cm}
\centerline{{\includegraphics[width=3.1cm]{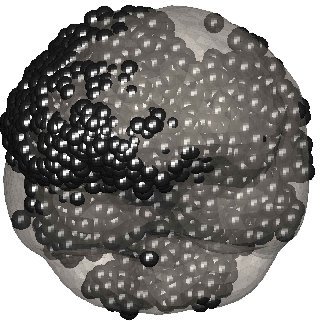}}}
\end{minipage}
\begin{minipage}{4cm}
\centerline{{\includegraphics[width=3.7cm]{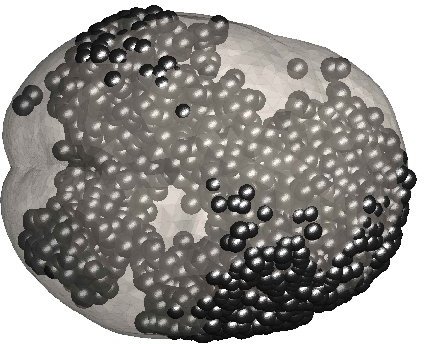}}}
\end{minipage}}
\caption{\label{fig:estimates theta1}Sagittal, coronal, and axial
projections (left to right) of the MAP estimate from EEG data of the
variance of the prior for the gamma hypermodel (top row) and for the
inverse gamma hypermodel (bottom row). In the top row, location of
the components of $\theta$ larger than 5\% of the maximum value are
marked with a bubble. In the bottom row only elements whose values
were greater than or equal to  85\% of the maximum were plotted,
indicating that the estimated variance is nearly maximal over the
entire brain.}
\end{figure}
\begin{figure}[t]
\centerline{\begin{minipage}{4cm}
\centerline{{\includegraphics[width=3.7cm]{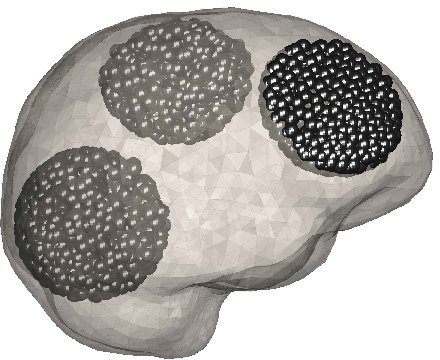}}}
\end{minipage}
\begin{minipage}{4cm}
\centerline{{\includegraphics[width=3.1cm]{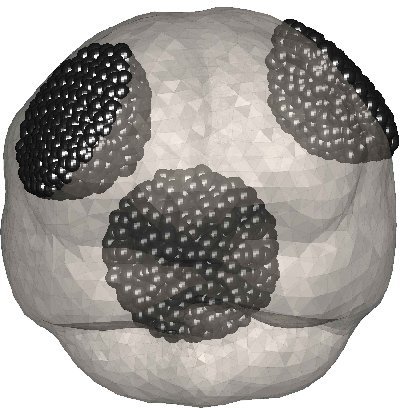}}}
\end{minipage}
\begin{minipage}{4cm}
\centerline{{\includegraphics[width=3.7cm]{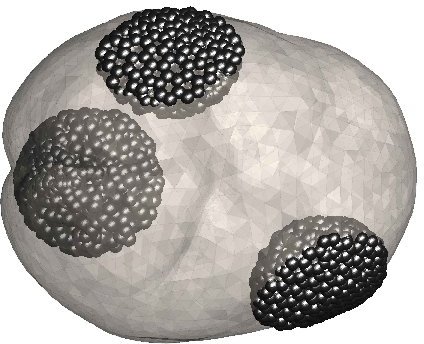}}}
\end{minipage}}
\centerline{\begin{minipage}{4cm}
\centerline{{\includegraphics[width=3.7cm]{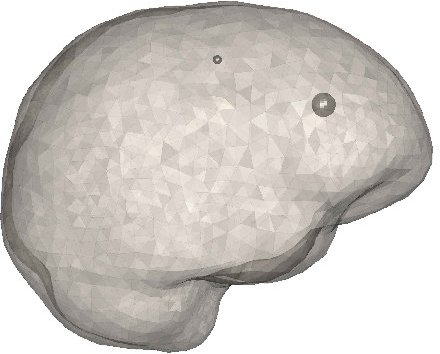}}}
\end{minipage}
\begin{minipage}{4cm}
\centerline{{\includegraphics[width=3.1cm]{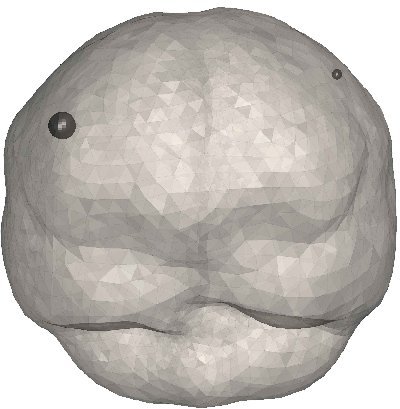}}}
\end{minipage}
\begin{minipage}{4cm}
\centerline{{\includegraphics[width=3.7cm]{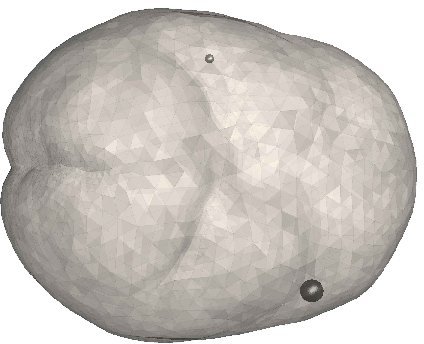}}}
\end{minipage}}
\caption{\label{fig:estimates theta2}Sagittal, coronal, and axial
projections (left to right) of the CM estimate from EEG data of the
variance of the prior for the gamma hypermodel (top row) and for the
inverse gamma hypermodel (bottom row). The thresholding level is set
to 5\% in this plot.}
\end{figure}
\begin{figure}[t]
\centerline{\begin{minipage}{4cm}
\centerline{{\includegraphics[width=3.7cm]{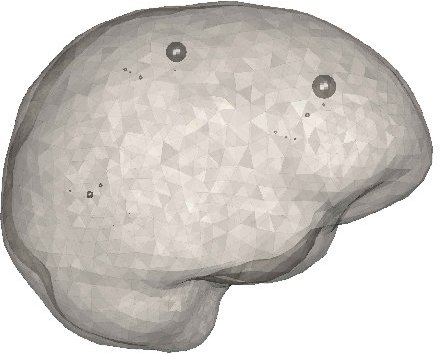}}}
\end{minipage}
\begin{minipage}{4cm}
\centerline{{\includegraphics[width=3.1cm]{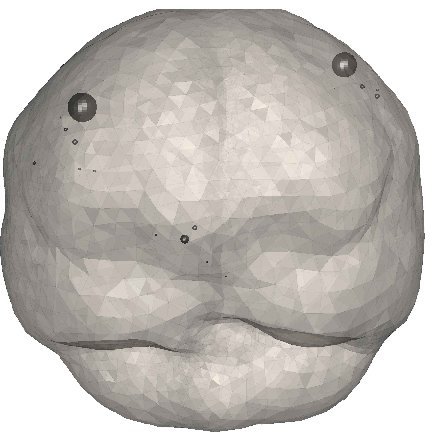}}}
\end{minipage}
\begin{minipage}{4cm}
\centerline{{\includegraphics[width=3.7cm]{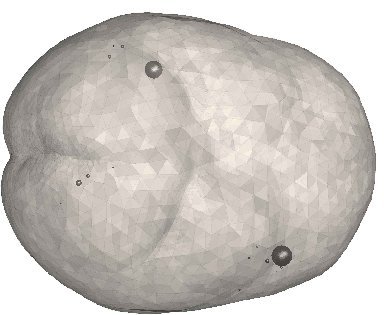}}}
\end{minipage}}
\centerline{\begin{minipage}{4cm}
\centerline{{\includegraphics[width=3.7cm]{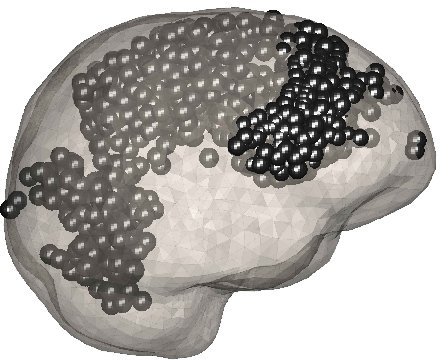}}}
\end{minipage}
\begin{minipage}{4cm}
\centerline{{\includegraphics[width=3.1cm]{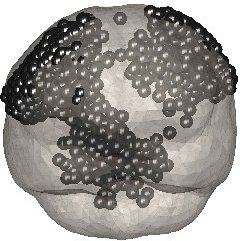}}}
\end{minipage}
\begin{minipage}{4cm}
\centerline{{\includegraphics[width=3.7cm]{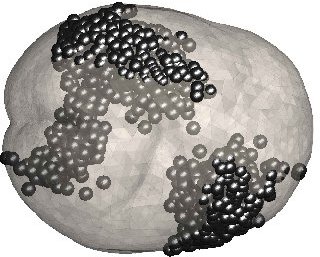}}}
\end{minipage}}
\caption{\label{fig:estimates theta3}Sagittal, coronal, and axial
projections (left to right) of the MAP estimate from MEG data of the
variance of the prior for the gamma hypermodel (top row) and inverse
gamma hypermodel (bottom row). The thresholding in the visualization
is as in Figure~\ref{fig:estimates theta1}.}
\end{figure}
\begin{figure}[t]
\centerline{\begin{minipage}{4cm}
\centerline{{\includegraphics[width=3.7cm]{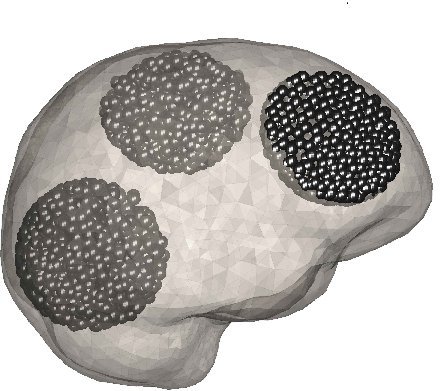}}}
\end{minipage}
\begin{minipage}{4cm}
\centerline{{\includegraphics[width=3.1cm]{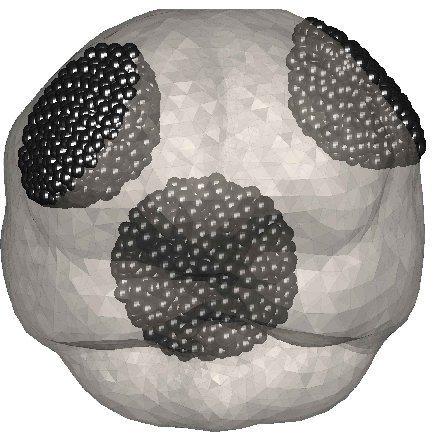}}}
\end{minipage}
\begin{minipage}{4cm}
\centerline{{\includegraphics[width=3.7cm]{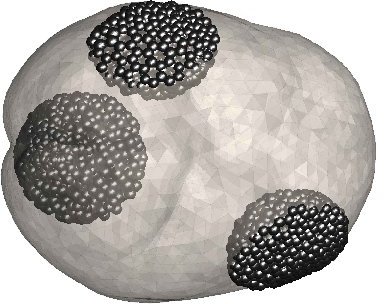}}}
\end{minipage}}
\centerline{\begin{minipage}{4cm}
\centerline{{\includegraphics[width=3.7cm]{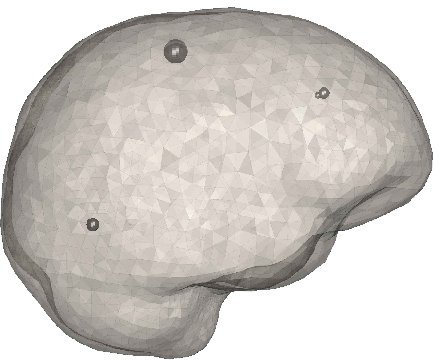}}}
\end{minipage}
\begin{minipage}{4cm}
\centerline{{\includegraphics[width=3.1cm]{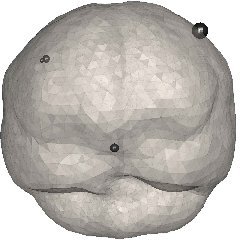}}}
\end{minipage}
\begin{minipage}{4cm}
\centerline{{\includegraphics[width=3.7cm]{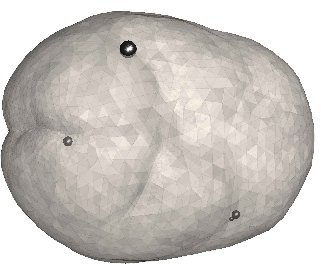}}}
\end{minipage}}
\caption{\label{fig:estimates theta4}Sagittal, coronal, and axial
projections (left to right) of the CM estimate from MEG data of the
variance of the prior for the gamma hypermodel (top row) and for the
inverse gamma hypermodel (bottom row). Here, the visualization
threshold is set to 5\%.}
\end{figure}
\begin{figure}[t]
\centerline{\begin{minipage}{4cm}
\centerline{{\includegraphics[width=3.7cm]{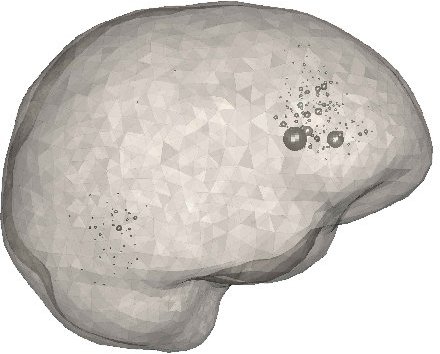}}}
\end{minipage}
\begin{minipage}{4cm}
\centerline{{\includegraphics[width=3.1cm]{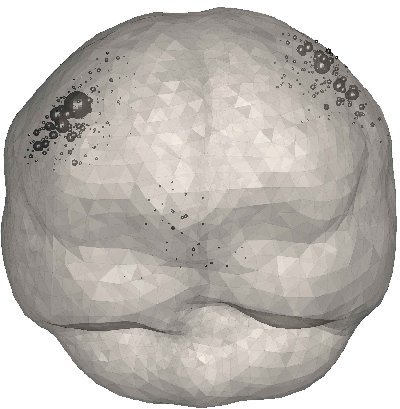}}}
\end{minipage}
\begin{minipage}{4cm}
\centerline{{\includegraphics[width=3.7cm]{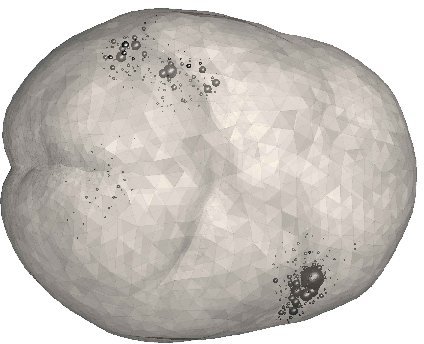}}}
\end{minipage}}
\centerline{\begin{minipage}{4cm}
\centerline{{\includegraphics[width=3.7cm]{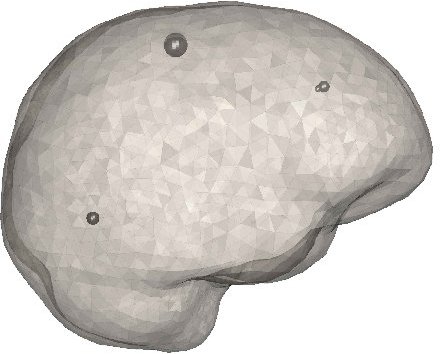}}}
\end{minipage}
\begin{minipage}{4cm}
\centerline{{\includegraphics[width=3.1cm]{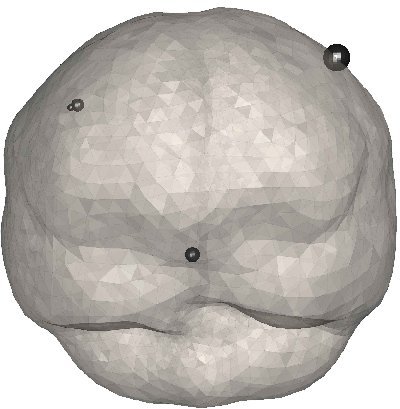}}}
\end{minipage}
\begin{minipage}{4cm}
\centerline{{\includegraphics[width=3.7cm]{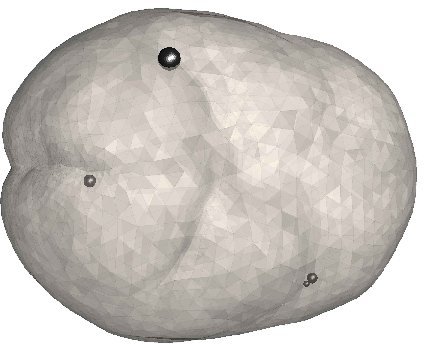}}}
\end{minipage}}
\caption{\label{fig:estimates postvar1}Sagittal, coronal, and axial
projections (left to right) of the sample based estimates of the
posterior variance from EEG data for the gamma hypermodel (top row)
and for the inverse gamma hypermodel (bottom row). The visualization
threshold is set at 5\% here.}
\end{figure}
\begin{figure}[t]
\centerline{\begin{minipage}{4cm}
\centerline{{\includegraphics[width=3.7cm]{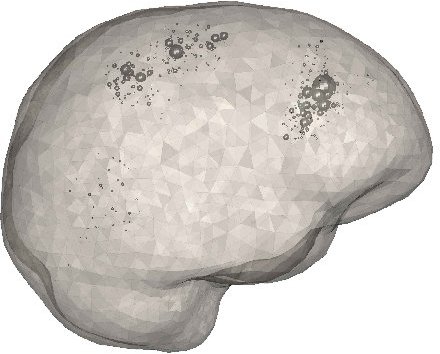}}}
\end{minipage}
\begin{minipage}{4cm}
\centerline{{\includegraphics[width=3.1cm]{variance_gamma_cm_meg_2}}}
\end{minipage}
\begin{minipage}{4cm}
\centerline{{\includegraphics[width=3.7cm]{variance_gamma_cm_meg_3}}}
\end{minipage}}
\centerline{\begin{minipage}{4cm}
\centerline{{\includegraphics[width=3.7cm]{variance_inversegamma_cm_meg_1}}}
\end{minipage}
\begin{minipage}{4cm}
\centerline{{\includegraphics[width=3.1cm]{variance_inversegamma_cm_meg_2}}}
\end{minipage}
\begin{minipage}{4cm}
\centerline{{\includegraphics[width=3.7cm]{variance_inversegamma_cm_meg_3}}}
\end{minipage}}
\caption{\label{fig:estimates postvar2}Sagittal, coronal, and axial
projections (left to right) of the sample based estimates of the
posterior variance from MEG data for the gamma hypermodel (top row)
and for the inverse gamma hypermodel (bottom row). The visualization
threshold is set to  5\%.}
\end{figure}
\begin{figure}[t]
\centerline{\begin{minipage}{5.5cm}
\centerline{{\includegraphics[width=4.5cm]{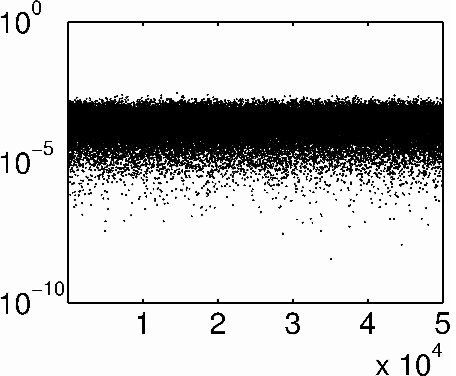}}}
\end{minipage} \hskip0.1cm
\begin{minipage}{5.5cm}
\centerline{{\includegraphics[width=4.41cm]{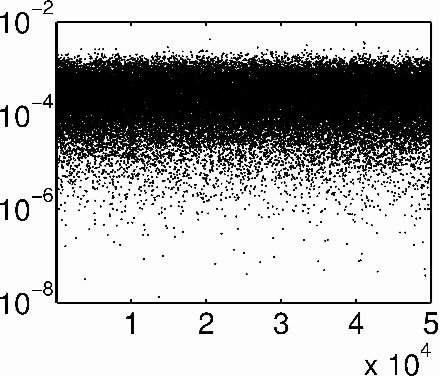}}}
\end{minipage}}
\centerline{\begin{minipage}{5.5cm}
\centerline{{\includegraphics[width=4.5cm]{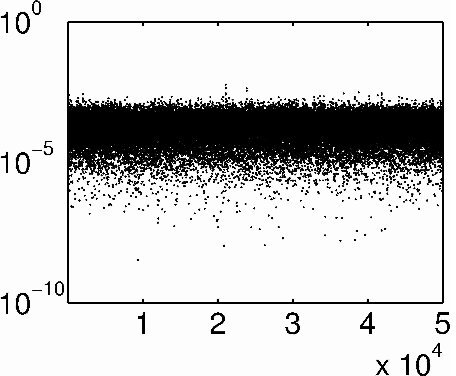}}}
\end{minipage}
\begin{minipage}{5.5cm}
\centerline{{\includegraphics[width=4.5cm]{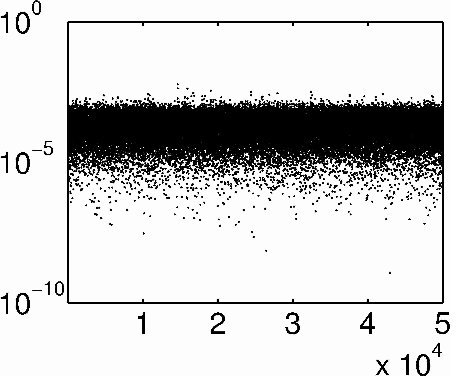}}}
\end{minipage}}
\caption{\label{fig:sample history 2} Sample histories of the
component corresponding to the maximum of the posterior mean
estimate of the current estimate from the EEG data and the gamma
hypermodel (top left) and with the inverse gamma hypermodel (bottom
left). In the right column we show the corresponding plots obtained
from the MEG data and the gamma distribution (top right) or the
inverse gamma distribution (bottom right).}
\end{figure}

To validate the hyperprior models in a more realistic geometry, we
performed EEG and MEG tests with a realistic human head model, based
on MRI data. We assume that both electric and magnetic fields are
measured outside the skull at 31 different locations, as shown in
Figure~\ref{fig:head geometry}. This model is partitioned into four
domains of constant electric conductivity: scalp, skull,
cerebrospinal fluid (CSF), and brain, as illustrated in
Figure~\ref{fig:head geometry}. Anisotropies in the brain as well as
possible electroconductive differences between the gray and the
white matter are ignored.

The electric and magnetic lead field matrices for this setup were
constructed using the complete electrode model and a Finite Element
Method (FEM), described in the Appendix, under the assumption that
all electrode contact impedances are equal to one. The FEM mesh was
generated in two stages. The meshes for the skull and the brain
domains were generated first using triangular elements. The meshes
for the scalp and CSF domains were then generated by positioning
prism elements, each to be subsequently divided into three
tetrahedral elements, between the brain and the skull surface. The
total number of tetrahedral elements in the head mesh is 108\,914.
The electric conductivities of the domains are given in Table
\ref{table: conductivities}. These values are as in \cite{wen}.

Lowest order $H(\hbox{div})$-conforming Raviart-Thomas elements were
applied for the source current density. In these elements, basis
functions are linear over the tetrahedron, vanish at one of the
vertices, and have a constant direction normal to the face opposite
to the vertex where they vanish \cite{braess}. For a curl-free
current density, e.g.\ a dipole source, it is necessary to use
$H(\hbox{div})$-conforming elements \cite{monk}, i.e.\ elements in
which basis functions and their divergences are square integrable.
Basis functions for $H(\hbox{div})$-conforming elements can be
interpreted to represent dipole-like currents \cite{tanzer}.

The electric potential $u$ was modelled using Lagrange elements
\cite{braess}. Quadratic Lagrange elements and a fifteen point Gauss
quadrature rule \cite{braess} were employed to generate simulated,
noiseless reference data, while the exploration of the posterior was
done using linear Lagrange elements and a four point Gauss
quadrature to speed up the computation. The resulting forward
modelling error for the electric lead field matrix in the brain
domain was approximately 3 \% in the Frobenius norm \cite{braess},
and approximately 2 \% for the magnetic lead field matrix.

The reference current density used to generate the reference data,
shown in Figure~\ref{fig: reference}, corresponds to the case where
we have three dipole-like source currents, one positioned deeply,
approximately 2.5 cm under the occipital lobe, and the other two on
the surface layer modelling the cortex. The source on the right of
the frontal lobe is almost tangential to the surface, and the source
close to the left central sulcus is almost normal to the surface.

As in the planar geometry example, the standard deviation in the
likelihood model is assumed to be 5\% of the maximum noiseless
signal. In this case, to simulate real situations, the noise is
added to the generated data.

The MAP estimate with the IAS algorithm and posterior mean by MCMC
sampling over the ROI are computed, corresponding to two different
hyperprior models (gamma, inverse gamma) and to the EEG and MEG
recording modalities. In these examples, the scaling and shape
parameter values are held constant at $\theta_0=10^{-7}$ and $\beta
= 1.55$ and up to 20 iterations of the IAS methods were allowed to
compute the MAP estimates. The ROI consisted of three disjoint sets
containing together $5\,982$ elements, see Figure~\ref{fig:
reference}. Corresponding to each combination of hyperprior model
and recording modality, an MCMC sample of size $M=50\,000$ is
generated, assuming that impressed currents differ from zero only
inside the ROI.

In Figures~\ref{fig:estimates alpha1} -- \ref{fig:estimates alpha2},
the MAP and posterior mean estimates of the current vector $\alpha$
using the EEG data are superimposed with three different projections
(sagittal, coronal, and axial) of the brain. The top row corresponds
to the gamma hyperprior and the bottom row to inverse gamma.
Interestingly, the MAP estimate with the gamma distribution is more
focal than with the inverse gamma, while with the posterior mean,
the inverse gamma yields more focal estimates. The same behavior is
also seen in the estimates based on the MEG data, see
Figures~\ref{fig:estimates alpha3} -- \ref{fig:estimates alpha4}.

Although focal, the estimate obtained from the EEG data is unable to
locate all the sources. The preference for finding the right
cortical source while missing the left cortical source may be
related to the orientation of the dipoles and the positioning of the
electrodes in the simulation.

Unlike in the case of the planar geometry, when using a realistic
head model the deep source in the MAP estimates is not driven to the
surface, indicating that there is enough geometric information in
the three dimensional positioning of the electrodes or magnetometers
to resolve the depth.

The MAP and posterior mean estimates of the variance vector $\theta$
of the prior are shown in Figures~\ref{fig:estimates theta1} --
\ref{fig:estimates theta2} using the EEG data and in
Figures~\ref{fig:estimates theta3} -- \ref{fig:estimates theta4} for
the MEG data. The results are in full agreement with the
corresponding estimates for the currents.

Finally, based on the MCMC sampling, we estimate the posterior
variances $\hbox{Var}(\alpha)$ of the current vector. The estimated
variances are visualized in Figures~\ref{fig:estimates postvar1} --
\ref{fig:estimates postvar2}. Interestingly, even when the posterior
estimates are blurry, the posterior variances are relatively focal,
giving additional information of the likely positions of the
sources.

To monitor the performance of the MCMC sampler, in
Figure~\ref{fig:sample history 2} we have plotted the sample
histories of the components of $\alpha$ and $\theta$ corresponding
to the index of maximum value of the conditional mean estimate. In
the realistic geometry, we do not encounter effects that would
suggest multimodality of the distribution, and the mixing and
convergence seem, by visual inspection, satisfactory.

\section{Discussion}

In this article, we consider the MEG/EEG inverse problems of
localizing few focal sources in the framework of Bayesian
hierarchical models. It is shown that by using a conditionally
Gaussian prior combined with generalized gamma distributions as
hyperpriors, a rich family of posterior densities ensues, and it is
possible to generate numerous estimates, some of which are closely
related to previously proposed, regularization based estimators.

We propose a simple and effective numerical algorithm, the Iterative
Alternating Sequential (IAS) algorithm for computing the MAP
estimate simultaneously for the current density and its variance.
The versatility of this approach is confirmed by its ability of
producing an efficient fixed point implementation of several well
known focal reconstruction methods based on non-quadratic penalties
or non-Gaussian prior distributions. Particular instances, that
correspond to a choice of few scalar parameters in the hypermodel,
include the minimum current estimate, the $\ell^p$-regularized
estimate and the limited support functional estimate. Furthermore,
the efficient numerical implementation using iterative solvers gives
also a natural interpretation in the statistical framework for the
FOCUSS algorithm when applied to the MEG/EEG imaging problem.
Compared to the empirical Bayes methods proposed in the literature
such as evidence maximization, or Automatic Relevance Determination
(ARD), Expectation Maximization (EM) and variational methods
\cite{sato,wipf}, the IAS algorithm is very fast, easy to implement
and does not require sophisticated minimization methods, nor does it
lead to intractable integral expressions. As shown in this article,
explicit expressions for the iterative minimization steps can be
found with numerous choices of the hyperprior parameters without
requiring conjugacy property of the hyperprior.

The different choices of the hypermodel parameter lead to algorithms
that behave qualitatively similarly with respect to the MAP
estimation: when the depth resolution due to the measurement
geometry is poor, as is the case in the local half space model,
superficial focal sources are localized well, while deep sources are
biased towards the surface. The MCMC analysis of the posterior
distributions, however, reveals a qualitative difference between the
hypermodels with different values of the hypermodel parameter $r$.
In the case where $r=1$, corresponding to the gamma hypermodel, the
MAP estimate coincides with the minimum current estimate and, like
the posterior mean estimate, is biased towards superficial sources.
With $r=-1$, yielding the inverse gamma hypermodel, the MAP estimate
is also biased towards superficial sources, but the posterior mean
is not. This qualitative difference may be interpreted by saying
that the smaller the parameter $r$, the less Gaussian the model, and
in the limit as $r\to\infty$ we obtain the minimum norm solution.
Note that for Gaussian distributions, the posterior mean and the MAP
estimates coincide.

From the point of view of Bayesian modelling paradigm, the
qualitatively different behavior of solutions corresponding to
different hyperpriors may seem strange, since in none of the cases
any a priori preference to superficial sources is given. The
explanation is related to the parameter values in the hyperpriors.
The scaling parameter $\theta_0$ in the gamma distribution was
chosen very small in our example with planar geometry to obtain good
localization in the tangential direction. Since the mean of the
gamma distribution is $\theta_0\beta$, this choice of $\theta_0$
favors small current dipoles in the conditional mean, and the
energetically easiest way to achieve this is to place all the
currents on the surface. The mean of the inverse gamma distribution,
$\theta_0/(\beta-1)$, is also a small number for this choice of
$\theta_0$, but since this distribution allows significantly larger
outliers, it has no difficulty in letting few large dipoles in the
lower layers explain the data. For completeness, we also performed
MCMC runs with the gamma distribution with considerably larger
scaling parameter values. The result suggest that when the scaling
parameter is large enough to allow  dipoles of the correct size in
the lower layers where the true dipole lies, the focal properties of
the posterior mean are lost. We conclude that the findings are in
line with the Bayesian philosophy and seem to suggest that the
inverse gamma prior is more suitable for deep sources.

When using a three-dimensional more realistic geometry, the question
of depth resolution becomes more delicate since the geometry starts
to play a significant role, and measurements from different
directions possibly contribute to the depth resolution more than the
prior.

The computed results obtained with the realistic head model are in
good agreement with the planar case results, suggesting that the
posterior mean estimate is most effective in combination with the
inverse gamma prior. The MAP estimate, on the other hand, is most
effective when applied in connection with the gamma prior. In
general, for the inverse gamma hypermodel the posterior mean
estimation produced better results with larger values of the scaling
parameter $\theta_0$ than MAP estimation, while for the gamma
hypermodel, MAP estimates were superior to CM estimates with larger
$\theta_0$ values. In these model both estimate types were found to
be more sensitive to the choice of the shape parameter $\beta$ than
in the case of the planar geometry, with large value of $\beta$
leading to blurred estimates and ultimately to invisibility of the
deep source current. The hyperparameter values $\theta_0 = 10^{-7}$
and $\beta = 1.55$, used in the realistic case, were chosen based on
visual inspection.

Future extensions of this work include a hierarchical extension of
the model where the values of the hypermodel parameters will be
chosen based on the data, and the extension of the formalism to
include time dependent sources with a longitudinal correlation
structure.

\section{Appendix: Complete electrode model and FEM for EEG/MEG}

This appendix describes briefly how the electric and magnetic lead
field matrices can be constructed through the complete electrode
model \cite{somersalo} and the FEM for a realistic three dimensional
head, denoted here by $\Omega$.

The complete electrode model assumes that a set of contact
electrodes $e_1, e_2,$ $\ldots, e_{L}$ with contact impedances $z_1,
z_2,$$ \ldots, z_{L}$ is attached to the boundary $\partial \Omega$.
The electrode potential values are collected into a vector $U =
(U_1,$$ U_2, \ldots,$$ U_L)$ and the electric potential field $u$
satisfies the equation
\begin{equation}
\label{neumann2} \nabla \cdot( \sigma \nabla u) = \nabla \cdot {\bf
J}, \quad \hbox{in}\, \, \Omega,
\end{equation}
as well as the boundary conditions
\begin{equation} \label{reunaehdot_loppu} \sigma \frac{\partial u}{\partial
n}\Big|_{\partial \Omega \setminus \cup_\ell e_\ell}  =  0, \quad
\int_{e_\ell} \sigma \frac{ \partial u}{\partial n} \, d s = 0,
\quad \Big( u + z_\ell \sigma \frac{\partial u}{\partial n} \Big)
\Big|_{e_\ell} =  U_\ell,
\end{equation}
with $\ell = 1,2, \ldots, L$. Additionally, the Kirchoff's voltage
law $\sum_{\ell = 1}^L U_\ell =  0$ is assumed to hold. The weak
form of (\ref{neumann2}) and (\ref{reunaehdot_loppu}) can be
formulated by requiring that $u \in H^1(\Omega)   =   \{ \, w \in
L^2(\Omega) \, : \, {\partial w}/{\partial r_i } \in L^2(\Omega), \,
i = 1,2,3 \, \}$ and ${\bf J} \in {\bf H}(\hbox{{div}; } \Omega) =
\{\, {\bf w} \in L^2(\Omega)^3 \, : \, \nabla \cdot {\bf w} \in
L^2(\Omega) \, \}$. These function spaces are thoroughly discussed
 e.g.\ in \cite{monk}.

The finite element discretized fields corresponding to $u \in
H^1(\Omega)$ and ${\bf J}^p \in H(\hbox{div;} \Omega)$ can be
defined as $u_\mathcal{T} = \sum_{i = 1}^{N_u} \zeta_i \psi_i$ and
${\bf J}_\mathcal{T} = \sum_{i = 1}^{N_J} \alpha_{i} \, {\bf
w}_{i}$, respectively.  Here, the functions $\psi_1, \psi_2, \ldots,
\psi_{N_u} \in H^1(\Omega)$ and ${\bf w}_{1}, {\bf w}_{2},
\ldots,{\bf w}_{N_J} \in H(\hbox{div};\Omega)$ are, respectively,
scalar and vector valued finite element basis functions, defined on
a shape regular finite element mesh $\mathcal{T}$ \cite{braess}, and
the coefficients form the coordinate vectors ${\zeta} =
(\zeta_1,\zeta_2$,$\ldots,\zeta_{N_u})^T$ and ${ \alpha} =
(\alpha_1,\alpha_2$,$\ldots,\alpha_{N_J})^T$. Furthermore, since the
sum of the electrode potentials is assumed to be zero in the
complete electrode model, it is required that ${ U} =
(U_1,$$U_2,$$\ldots,$$ U_L)^T = {R} \tilde{\zeta}$,
 where $\tilde{\zeta}$ is a vector and ${ R} \in \mathbb{R}^{L \times (L-1)}$ is a matrix with entries
 $R_{1,j} = -R_{j+1,j} = 1$ for $j
 = 1, 2, \ldots,L-1$, and  otherwise $R_{i,j} = 0$.

The vectors $\alpha, \zeta$ and $\tilde{{\zeta}}$ are linked through
a symmetric and positive definite linear system of the form
 \begin{equation}
\label{u_system} \left[ \begin{array}{cc} { B} & { C} \\
{ C}^T & { G}
\end{array} \right] \left[ \begin{array}{c} {\zeta}  \\
\tilde{{\zeta}}
\end{array} \right] = \left[ \begin{array}{c} { F} {\alpha }  \\
{ 0}
\end{array} \right],
\end{equation} in which the submatrix entries are given by
\begin{eqnarray*}
\label{matrix_M} F_{i,k} & = & \int_\Omega (\nabla \cdot {\bf w}_k)
\psi_i \, d
\Omega,  \\
B_{i, j}  & = & \int_{\Omega} \sigma \nabla \psi_i \cdot \nabla
\psi_j \, d \Omega + \sum_{\ell = 1}^L \frac{1}{z_\ell}
\int_{e_\ell}
\varphi_i \varphi_j \, dS, \\
C_{i, j}  & = &  - \frac{1}{z_1} \int_{e_1} \varphi_i \, dS +
\frac{1}{z_{j+1}} \int_{e_{j+1}}
\varphi_i \, dS, \\
G_{i, j}  & = & \frac{1}{z_j} \int_{e_j} \,  dS +
\frac{\delta_{i,j}}{z_{j+1}} \int_{e_{j+1}} d S, \label{matrix_G}
\end{eqnarray*}
where $\delta_{i,j}$ denotes the Kronecker delta. The system
(\ref{u_system}) arises from the Ritz-Galerkin discretization
\cite{braess} of the weak form of (\ref{neumann2}) and
(\ref{reunaehdot_loppu}). Similarly, a discretized version of the
Biot-Savart law (\ref{biot savart}) can be expressed as $B = W
\alpha - V \zeta$, where $B$ is a vector containing the magnetic
field values at the measurement locations, and the matrices are
defined by
\begin{eqnarray*}
W_{i,3(j-1)+k}  & = &  \int_\Omega \frac{{\bf e}_k \cdot {\bf w}_j
\! \times \! ({\bf r}_i - {\bf r})}{| {\bf r}_i - {\bf r}|^3} \, d
{\bf r}, \quad \hbox{and} \\ V_{i,3(j-1)+k} & = &
 \int_\Omega \frac{{\bf e}_k \cdot \sigma \nabla \psi_j \! \times \! ({\bf r}_i - {\bf
r})}{| {\bf r}_i - {\bf r}|^3} \, d {\bf r},
\end{eqnarray*}
with ${\bf r}_j$ denoting the $j$th measurement location and ${\bf
e}_k$ denoting the $k$th natural basis vector.

The dependences of $U$ and $B$ on the vector $\alpha$ are described
by the electric and magnetic lead field matrices $M^{\rm e}$ and
$M^{\rm m}$, respectively. These matrices are given by
\begin{eqnarray*}
M^{\rm e} & = &  { R} ( { C}^T { B}^{-1} { C} - G )^{-1} { C }^T {
B}^{-1} { F}, \\
M^{\rm m} & = & W  - V ( B - C G^{-1} C^T)^{-1} F,
\end{eqnarray*}
as can be verified through straightforward linear algebra
manipulations. Note that these expressions are valid only if a set
of contact electrodes is attached to the head during the magnetic
field measurement.

\bibliographystyle{siam}
\bibliography{references}

\end{document}